\documentclass[12pt]{iopart}
\usepackage{iopams}
\usepackage{setstack}
\usepackage{graphicx}
\usepackage{harvard}
\usepackage{nicefrac}
\newcommand{\vect}[1]{\bi{#1}}

\newcommand{\bhsi}{\ensuremath{B_{\mathrm{HS}i}}}
\newcommand{\bhsf}{\ensuremath{B_{\mathrm{HS}f}}}
\newcommand{\bshi}{\ensuremath{B_{\mathrm{SH}i}}}
\newcommand{\bshf}{\ensuremath{B_{\mathrm{SH}f}}}

\begin{document}

\title[Via Hexagons to Squares in Ferrofluids]{Via Hexagons to Squares in
Ferrofluids:\\
Experiments on Hysteretic Surface Transformations
under Variation of the Normal Magnetic Field}
% from Hexagonal to square planforms in ferrofluids

\author{C Gollwitzer, I Rehberg, and R Richter}

\address{Experimentalphysik V
Universit\"atsstr. 30 95445 Bayreuth, Germany}
\ead{Christian.Gollwitzer@uni-bayreuth.de, Ingo.Rehberg@uni-bayreuth.de, Reinhard.Richter@uni-bayreuth.de}

\begin{abstract}
We report on different surface patterns on magnetic liquids following the Rosensweig
instability. We compare the bifurcation from the flat surface to a hexagonal array of spikes with the
transition to squares at higher fields. From a radioscopic mapping of the surface topography we
extract amplitudes and wavelengths.
For the hexagon--square transition, which is complex because of coexisting domains, 
we tailor a set of order parameters like peak--to--peak distance, circularity, angular correlation
function and pattern specific amplitudes from Fourier space. These measures enable us to quantify
the smooth hysteretic transition. Voronoi diagrams indicate a pinning of the domains. Thus
the smoothness of the transition is roughness on a small scale. 
\end{abstract}

\maketitle

\section{Introduction}
\begin{figure}[htb]
\centering
(a)\raisebox{0.05\linewidth}{\includegraphics[width=0.45\linewidth]{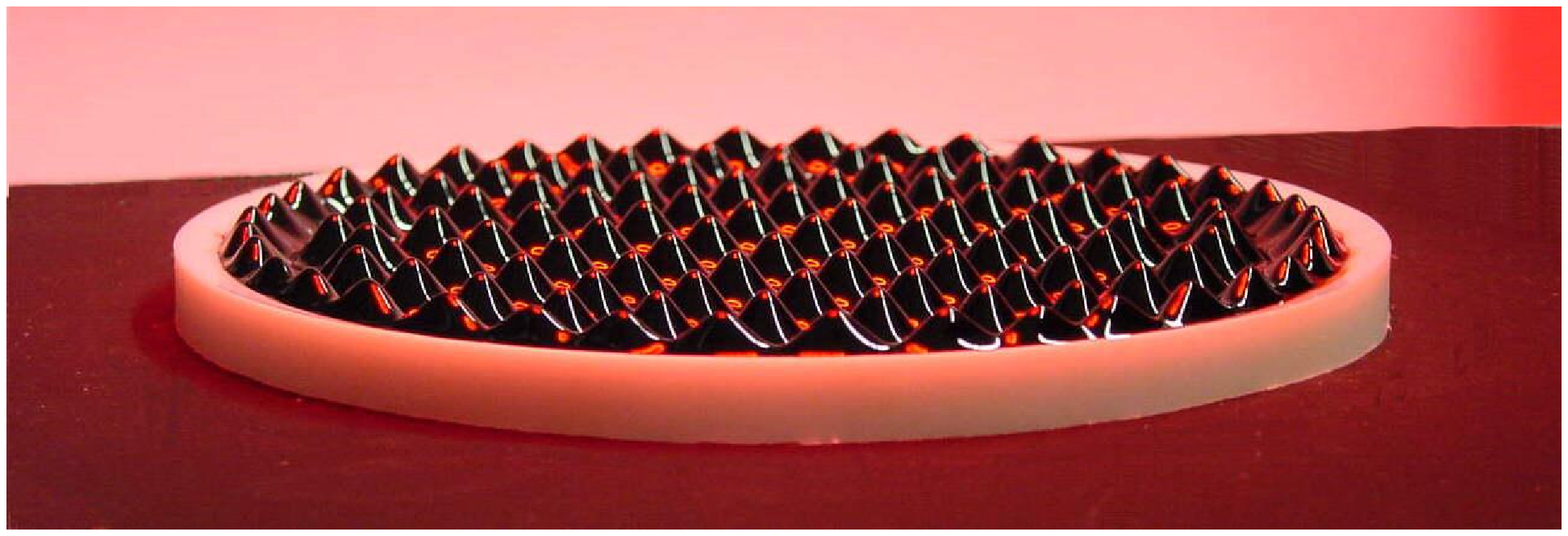}}
(b)\includegraphics[width=0.45\linewidth]{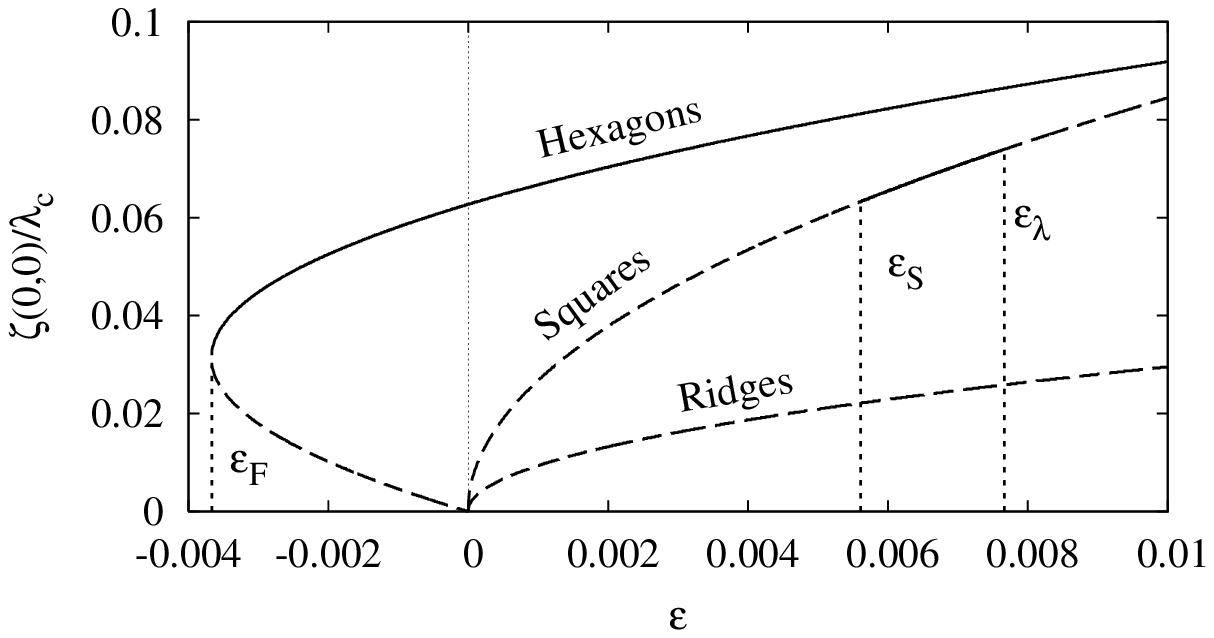}

\caption{The Rosensweig instability. Photo of the Rosensweig pattern~(a) and
bifurcation diagram~(b) following \citeasnoun{friedrichs2001}.} \label{fig:intro}
\end{figure}

The formation of static liquid mountains, floating on the free
surface of a magnetic fluid (MF), when subjected to a vertical
magnetic field is a fascinating phenomenon. It was uncovered by
\citeasnoun{cowley1967} soon after the synthesis of the first
ferrofluids and thus has served as a ``coat of arms'' for the field of
magnetic fluid research. The fascination stems in part from the fact
that liquid crests which persist without motion are not a familiar
experience (see Figure~\ref{fig:intro}\,a).

In contrast to pattern formation in dissipative systems
\cite{cross1993} the phenomenon can be described by an energy
functional, which comprises hydrostatic, magnetic and surface energy
\cite{gailitis1969,kuznetsov1976,gailitis1977}. As the surface profile deviates
from the flat reference state, the contributions of the hydrostatic
and the surface energy grow whereas the magnetic energy decreases.
For a sufficiently large magnetic induction $B>B_c$, this
gives rise to the normal field, or Rosensweig, instability. By
minimizing the energy functional an amplitude equation can be
derived, which has three roots \cite{friedrichs2001}, describing
liquid ridges, squares and hexagons, as sketched in
Figure~\ref{fig:intro}\,(b). The hexagons appear first due to a
transcritical bifurcation and their amplitude is given by
\begin{equation}
A_{H}=\frac{\gamma (1+\varepsilon) + \sqrt{\gamma^2
(1+\varepsilon)^2 + 4\,\epsilon g_\mathrm{h}}}{2 g_\mathrm{h}},
\label{eq:AE.hexagons}
\end{equation}
where $\varepsilon=(B^2-B_c^2)/B_c^2$ is the bifurcation parameter,
and $\gamma$, $g_\mathrm{h}$ are scaling parameters. The hysteretic
range $[\varepsilon_\mathrm{F}, 0]$ increases with increasing
susceptibility $\chi$.
Moreover we have a supercritical bifurcation
to squares, which become stable in the region
$\varepsilon_S<\varepsilon<\varepsilon_{\lambda}$. Their amplitude
is described by
\begin{equation}
A_{S}=\sqrt{\frac{\varepsilon}{g_\mathrm{s}}}.
\label{eq:AE.squares}
\end{equation}
%The ridges are also following a supercritical bifurcation
%\cite{zaitsev1969}, but are always unstable.
The bifurcation to ridges is also supercritical \cite{zaitsev1969}, but they are always unstable.
The parameters
$g_\mathrm{h}$,\,$g_\mathrm{s}$, and $\gamma$ depend on the wave
number $k$ and the susceptibility  $\chi$ and can be estimated for
rather small $\chi$ only \cite{friedrichs2001}.

The hexagon branch is {\it subcritical}, which complicates a
quantitative description of the Rosensweig instability, both in the
linear, as well as in the nonlinear approach.

A {\it linear description} of the Rosensweig instability is amenable
in theory, but restricted to small amplitudes. In experiments
however, small amplitudes are short-lived. 
Thus a new pulse technique
has been applied in order to measure the wave number
of maximal growth during the increase of the pattern \cite{lange2000}. Also the decay of metastable
patterns within the linear regime has been investigated in theory
and experiment by \citeasnoun{reimann2003}. Predictions for the
growth rate of the pattern amplitude by \citeasnoun{lange2001} are
tested by means of a novel magnetic detection technique
\cite{reimann2005}, which is capable to measure the pattern
amplitude with a high resolution in time (7k samples/sec). The achievementsin the linear regime are summarized by \citeasnoun{lange2006} 
and \citeasnoun{richter2006}.

A {\it nonlinear description} of the instability is, despite the
progress reported above, still restricted to small susceptibilities
and a linear magnetization law. Thus a full numerical approach to
the nonlinear problem, based on the finite element method (FEM), is
most welcome. Its achievements are reported in this issue
\cite{lavrova2006b}. From an experimental point of view, the final,
nonlinear state is difficult to access because the dark and steep
structures are an obstacle for standard optical measurement
techniques. Thus we developed a new approach to record the
full surface profiles, from which the bifurcation diagram can than
be established. The results obtained in the hexagonal regime are
summarized in the next section. The further parts of the article are
devoted to new investigations focusing on the transition from the
hexagonal to the square planform.

\section{Review of the Results in the Hexagonal Regime}
\label{sect:hexreview}
\begin{figure}[htb]
\centering
(a)\includegraphics[width=0.45\linewidth]{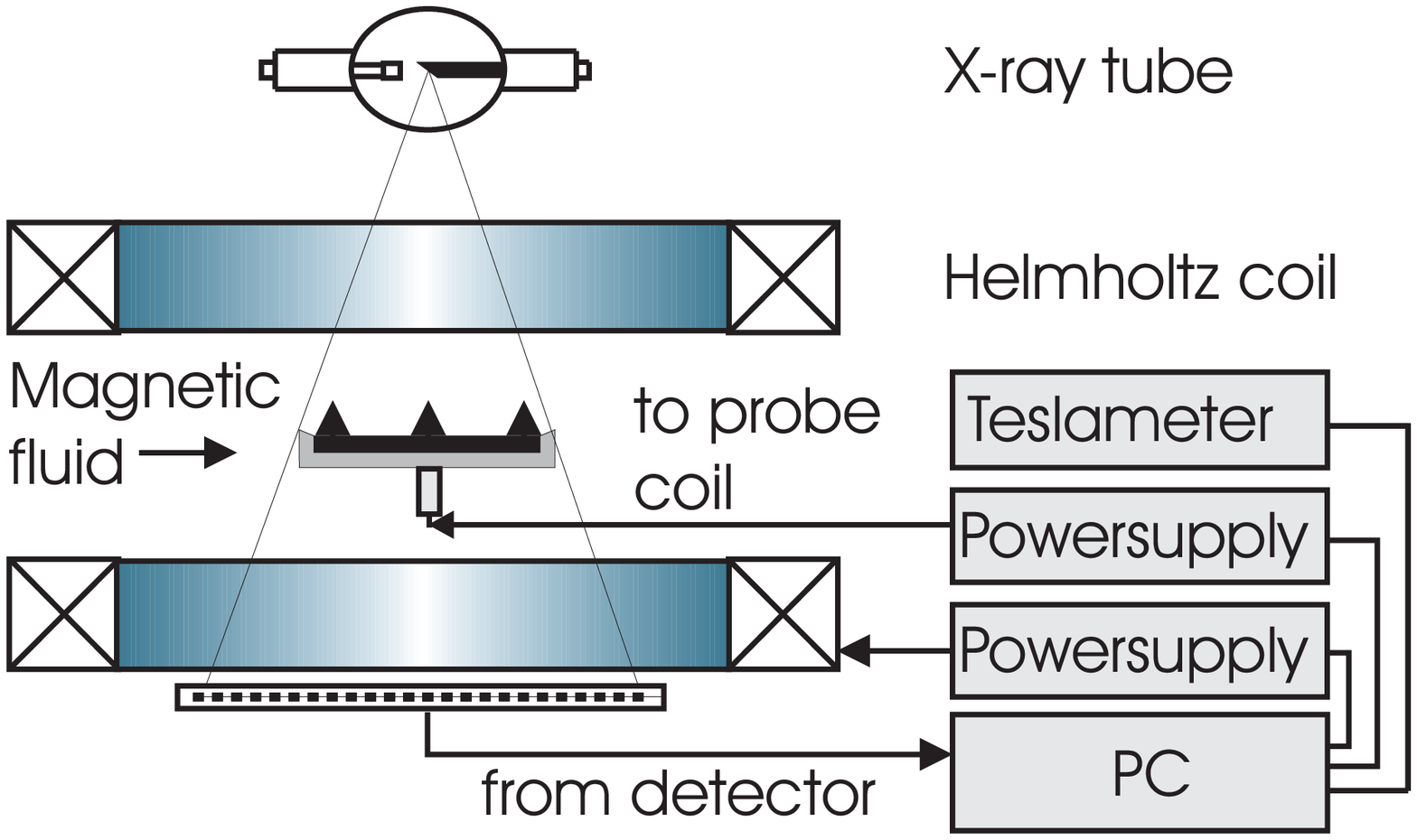}
(b)\raisebox{0.03\linewidth}{\includegraphics[width=0.45\linewidth]{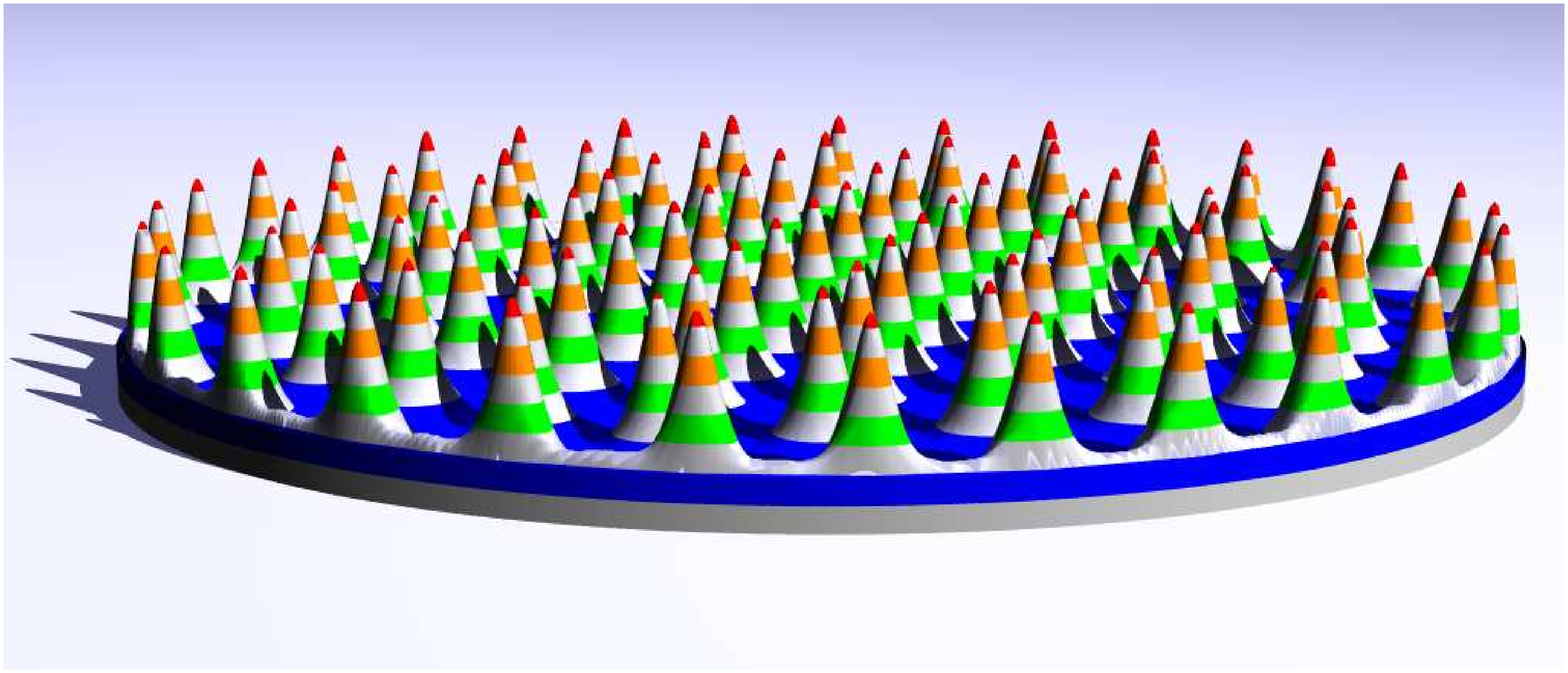}}

\caption{Measuring the surface relief via radioscopy: (a)
principle of the experimental setup, (b) reconstructed
surface relief. The radius of the vessel is $60\,$mm, its depth $3\,$mm.
The colour stripes have a width of $1\,$mm. The parameters of the fluid EMG\,901
(Ferrotec) are: $\chi=2.2$, $\rho=1.406
\rm\,g\,cm^{-3}$, and $\sigma=25 \pm 0.7 \rm\,mN/m$. Figures from
\citeasnoun{richter2005}} \label{fig:xray.principle}
\end{figure}

To overcome the hindrances to optical observation we utilize a
radioscopic technique which is capable to record the full surface
relief in the {\it centre} of
the vessel, far away from distortions by the edges
\cite{richter2001}. The experimental setup is sketched in
figure~\ref{fig:xray.principle}\,(a). A Teflon$^{\circledR}$ vessel is
filled with MF up to the brim and placed on the common axis midway
between a Helmholtz pair of coils. An X-ray tube is mounted above
the centre of the vessel at a distance of 1606\,mm. The radiation
transmitted through the fluid layer and the bottom of the vessel is
recorded by an X-ray sensitive photodiode array detector (16 bit).
The full surface relief, as presented in
figure~\ref{fig:xray.principle}\,(b) is then reconstructed from the
calibrated radioscopic images.

\begin{figure}[bh]
\includegraphics[width=\linewidth]{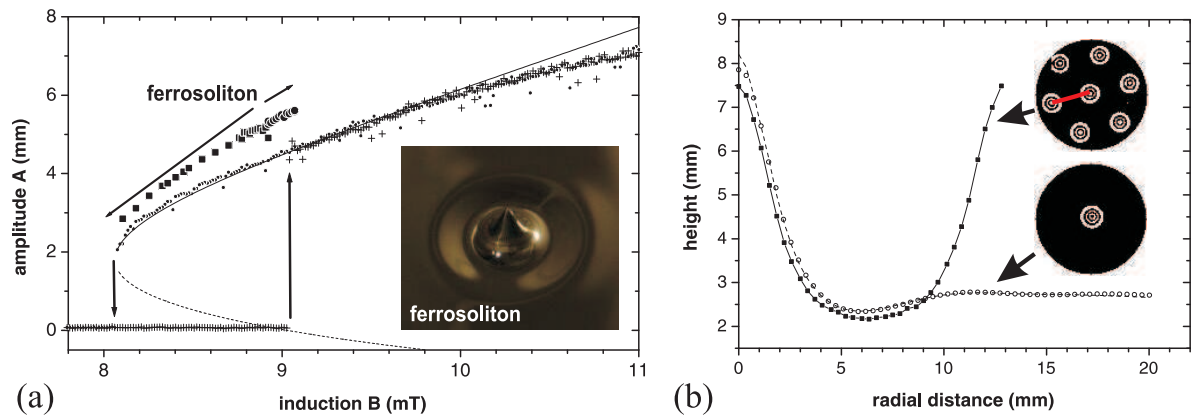}
\caption{ (a)~The amplitude of the pattern  versus the
magnetic induction. The crosses  (dots)  mark the values for
increasing (decreasing) induction, respectively. The solid (dashed)
lines display the least squares fit to~(\ref{eq:AE.hexagons})
with $\gamma=0.281$ and $g_\mathrm{h}=0.062$. The full circles
(squares) give the amplitude of the localized spike (see inset)
initiated at $B=8.91\, \rm mT$ for increasing (decreasing) induction
respectively. (b)~Height profiles for one period of the
hexagonal pattern (filled squares), and for two different solitons
(marked by open symbols and a dashed line). Figures from
\citeasnoun{richter2005}} \label{fig:soliton}
\end{figure}

With this set-up we measured the top-to-bottom amplitude $A$ of the
fluid pattern in the centre of the vessel.
Figure~\ref{fig:soliton}\,(a) displays the hysteretic behavior of
$A(B)$ for the adiabatic increase and decrease  of $B$, depicted by
crosses and dots, respectively. The solid and dashed lines display a
convincing fit to (\ref{eq:AE.hexagons}). We used a concentrated
MF ($\chi=2.2$) to realize a large hysteretic
regime, enabling us to investigate the stability of the flat surface
against local perturbations, within the bistable regime. For that
purpose a small cylindrical air coil was placed under the centre of the vessel,
as depicted in figure~\ref{fig:xray.principle}\,(a). This allows for a local
increase of the magnetic induction. A local pulse of
$B_+=0.68\,\mbox{\rm  mT}$ added to the uniform field of $B=8.91\,\rm
mT$ produces a single stationary spike of fluid, surrounded by a
circular dip, which does not disperse  after $B_+$ has been turned
off. The inset of figure~\ref{fig:soliton}\,(a) gives a picture of this
radially-symmetric state which will be referred to as a
\emph{ferrosoliton}. The range of stability of a soliton is marked in
figure~\ref{fig:soliton}\,(a) by full squares and full circles alongside 
the hysteretic regime.

The ferrosoliton is a stable non-decaying structure; it remains
intact for days -- for a movie see \citeasnoun{castelvecchi2005}. In
contrast to previously detected dissipative 2d-solitons, like
oscillons \cite{umbanhowar1996}, it is a static object without any
dissipation of energy. Thus the stabilization mechanism discussed
for dissipative 2d-solitons, a balance of dissipation versus
nonlinearity, can not be valid here. In order to shed some light on
its stabilization mechanism we compare in figure~\ref{fig:soliton}\,(b)
the height profile of solitons with the one of regular Rosensweig
peaks. The width of the soliton is equal to the period of the
lattice. This indicates that a spreading of the local
perturbation is locked by the periodicity of the Rosensweig lattice,
as suggested in a general context by \citeasnoun{pomeau1986}. This
wave-front locking could be verified in a conservative analogue of
the Swift--Hohenberg equation \cite{richter2005}.

\begin{figure}
\centering
\includegraphics[height=0.35\linewidth]{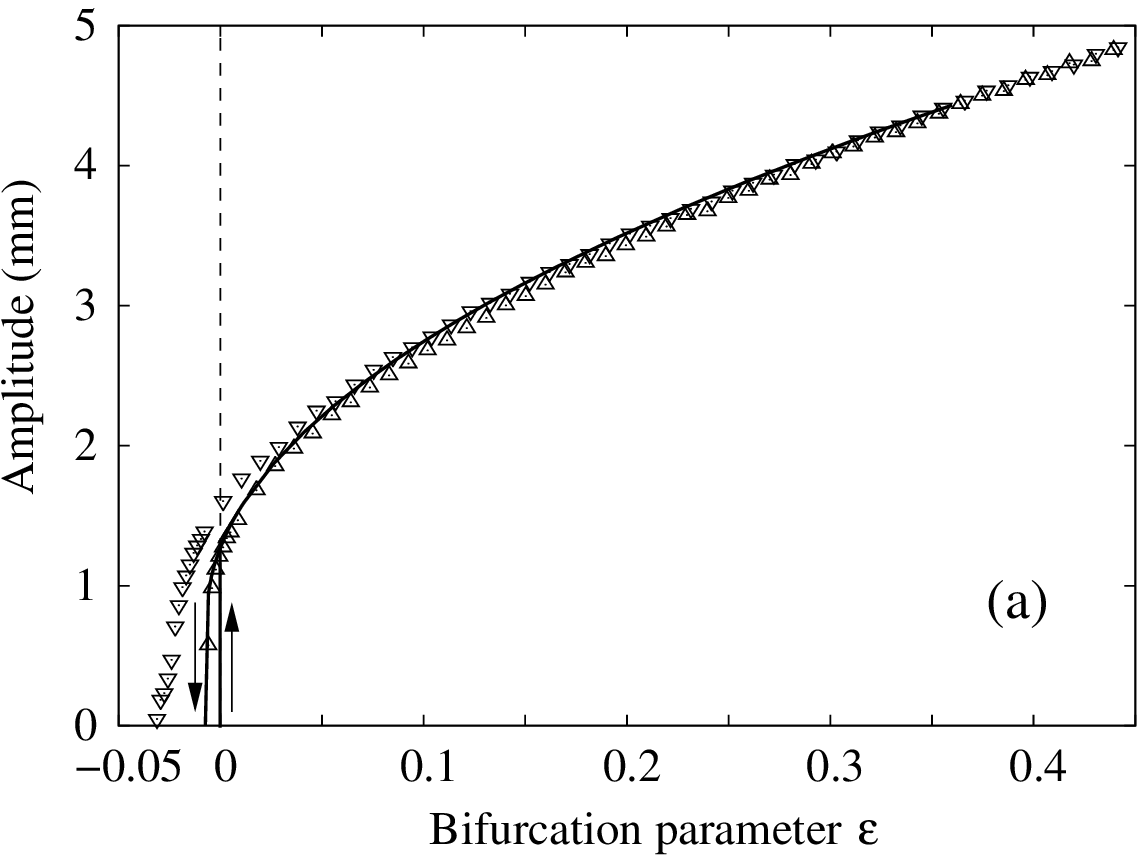}
\includegraphics[height=0.35\linewidth]{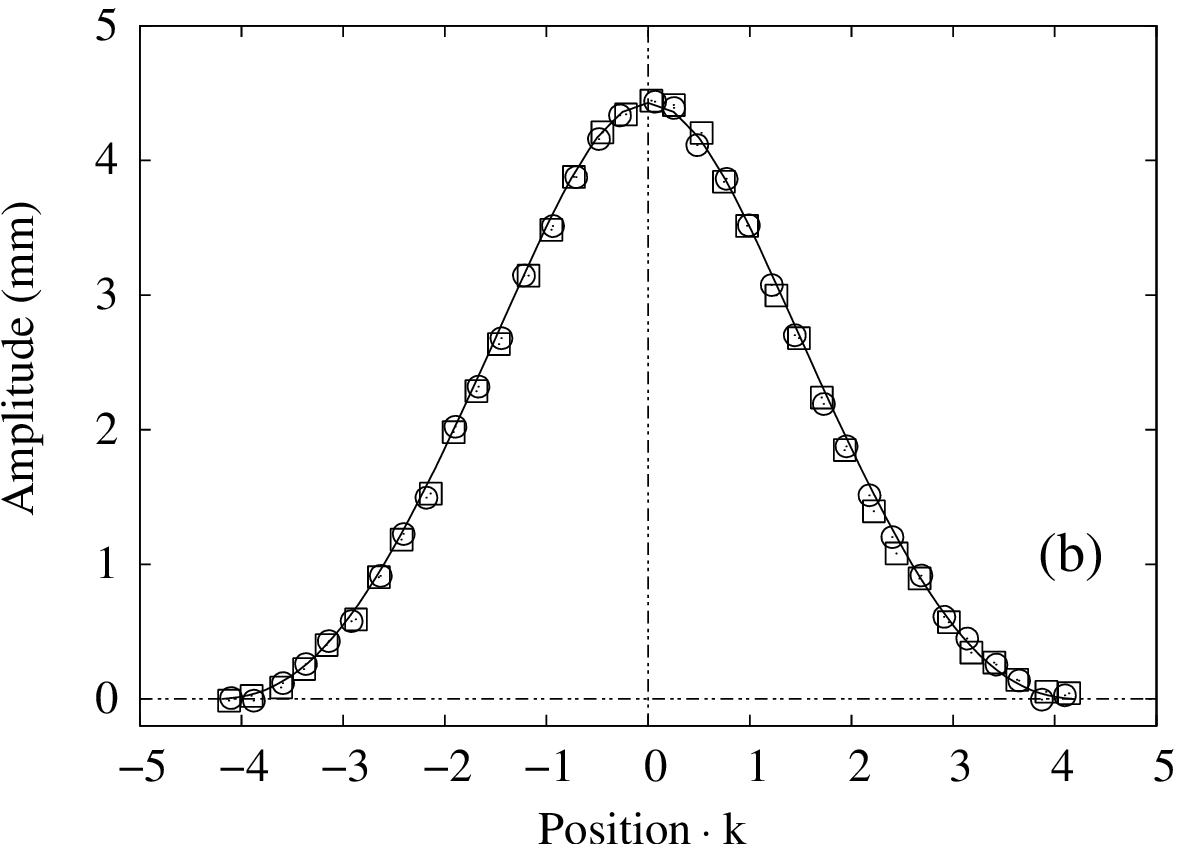}

\caption{Comparison of experimental and numerical results.
(a) The pattern amplitude for increasing~(decreasing) $\varepsilon$ is denoted by upward~(downward) triangles, respectively.
(b) The peak profiles at $\varepsilon=0.35$ for increasing~(decreasing) $\varepsilon$ are marked by squares~(circles), respectively.
The solid lines give the corresponding numerical results.
Figures from \citeasnoun{gollwitzer2006}.}
\label{fig:hex.comparison}
\end{figure}

From qualitatively new features we turn our attention now to a
\emph{quantitative} comparison of experimental and numerical
results. To make the comparison more feasible, we used a deeper container
($h=10\,$mm).  Its floor is 
flat within a diameter of $130\,$mm, outside of which it is inclined
upwards at $32$~degrees,
so that the thickness of the fluid layer smoothly decreases down to 
zero towards the side of the vessel. We select a MF available in large amounts,
which has a smaller susceptibility~($\chi=1.17$). 
Figure~\ref{fig:hex.comparison}\,(a) displays the evolution of the pattern
amplitude (triangles), which shows a reduced hysteresis in
comparison to figure~\ref{fig:soliton} due to the reduced $\chi$. The two solid lines give the
result of the FEM calculations taking into account the measured
fluid parameters and the nonlinear magnetization curve, as presented
in detail by \citeasnoun{gollwitzer2006}. Note, that the agreement
was obtained without any free fitting parameter.
Figure~\ref{fig:hex.comparison}\,(b) displays the measured and
calculated peak profiles, which match as well.

\section{The Hexagon--Square-Transition}
\begin{figure}
\centering
(a)\includegraphics[width=0.2\linewidth]{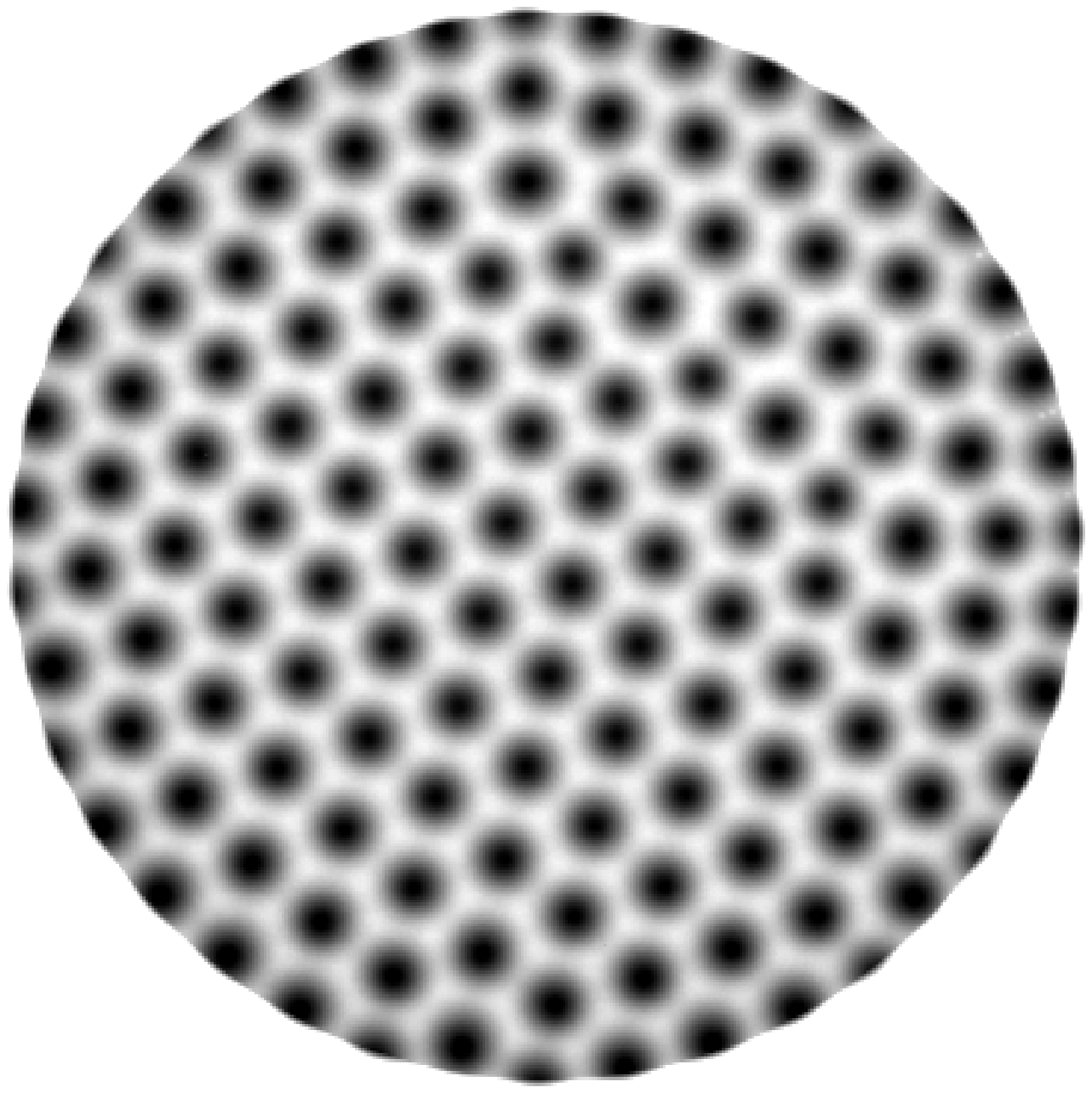}
(b)\includegraphics[width=0.2\linewidth]{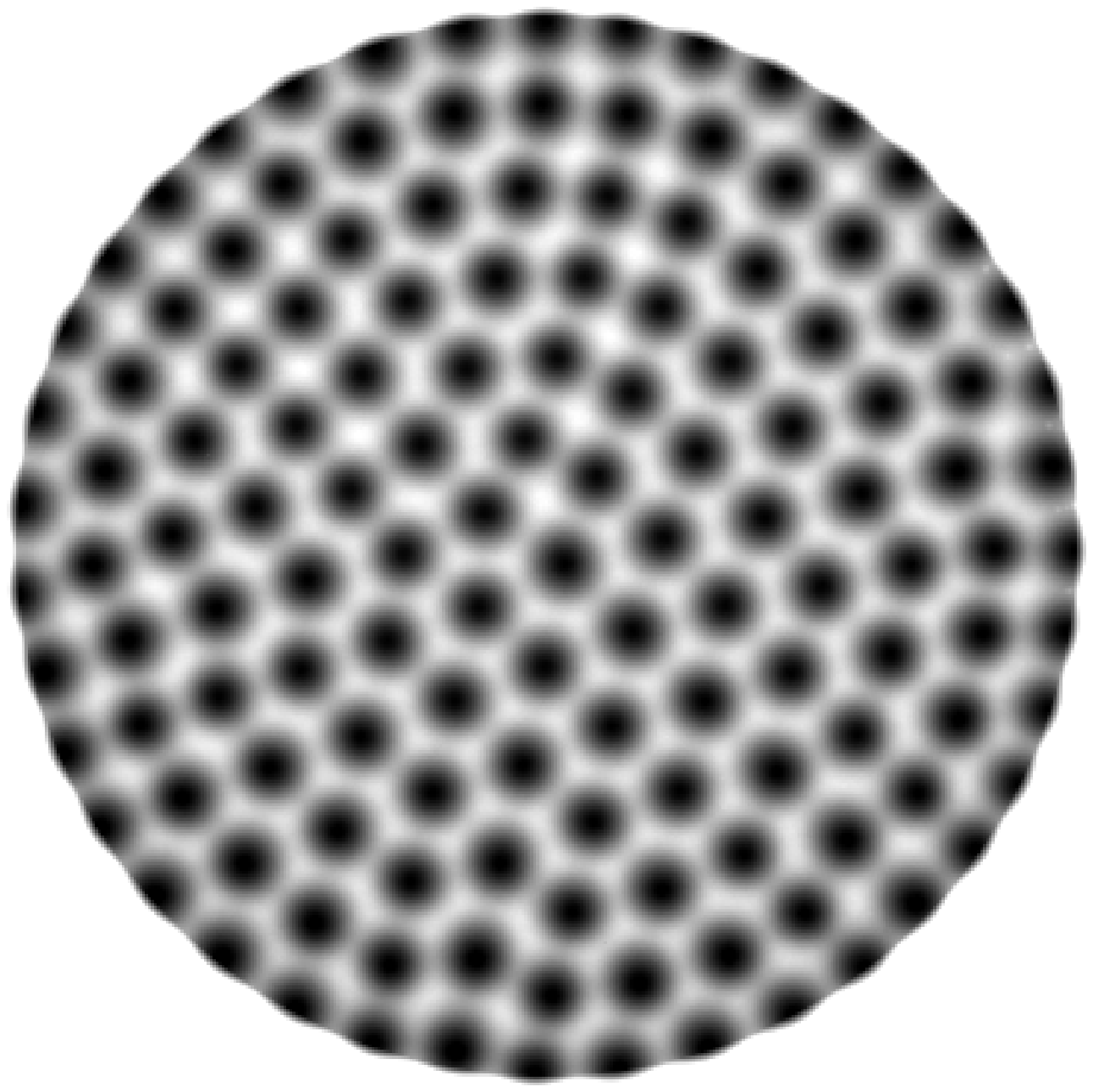}
(c)\includegraphics[width=0.2\linewidth]{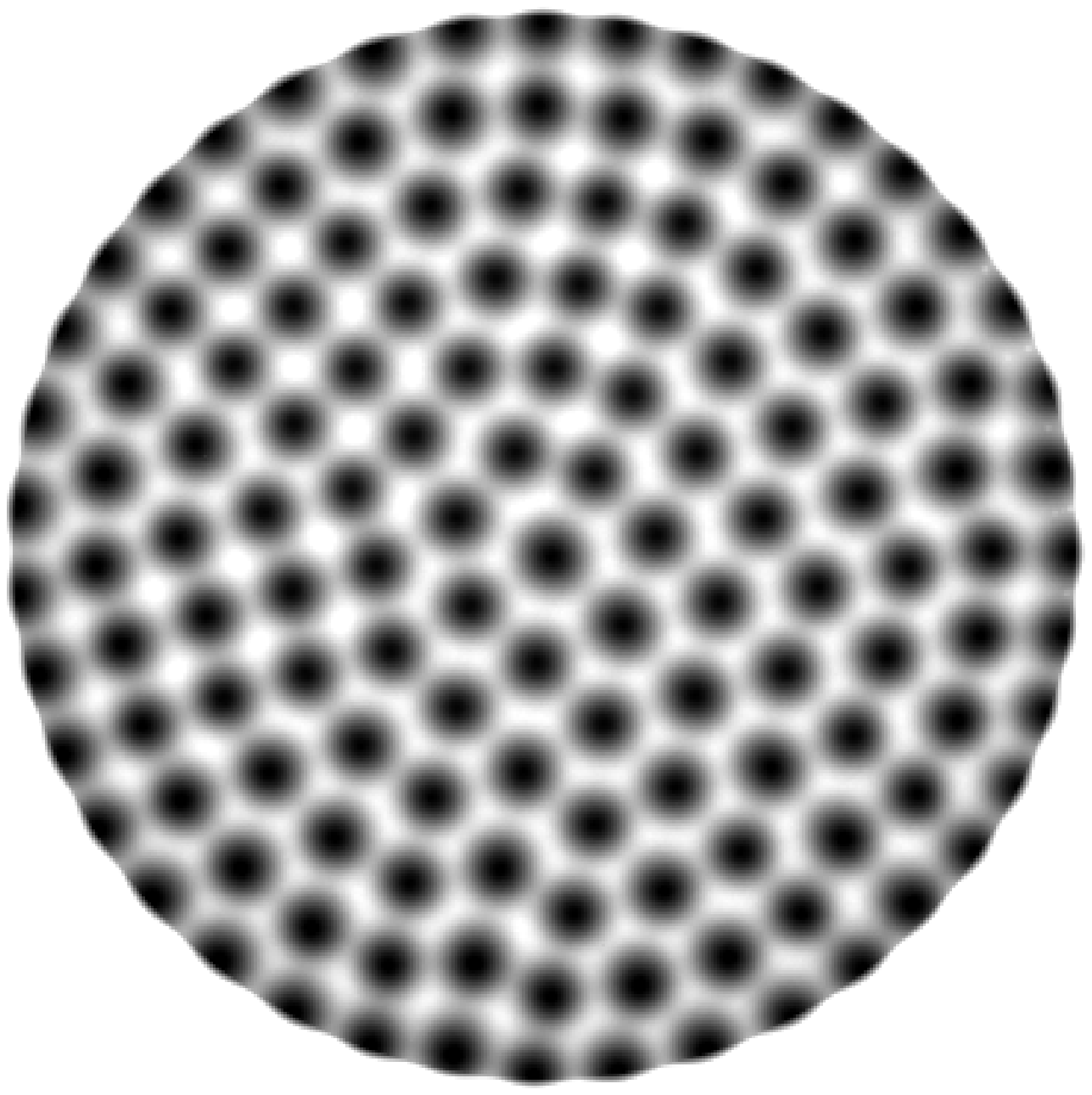}
(d)\includegraphics[width=0.2\linewidth]{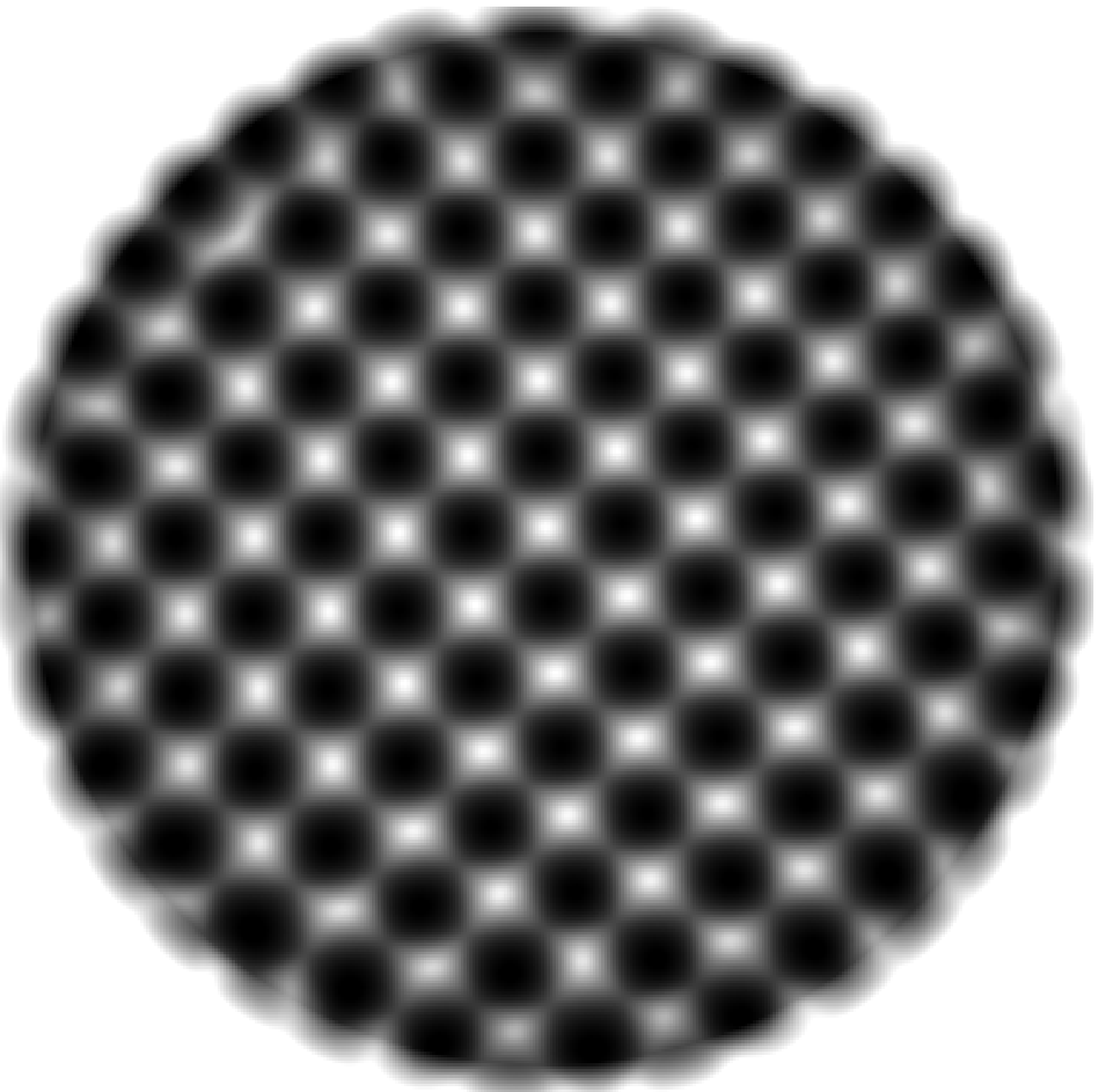}

\caption{X-ray absorption images of two different states of the pattern observed with the fluid
APG\,512\,A from Ferrotec. Fluid parameters are: initial susceptibility $\chi=1.17$, density
$\rho=1236\,\mathrm{kg\,m^{-3}}$, surface tension $\sigma=30.6\,\mathrm{mN\,m^{-1}}$, viscosity
$\eta=120\,\mathrm{mPa\,s}$. 
Hexagonal pattern~(a) observed at $B=19.78\,\mathrm{mT}$ and square pattern~(d)
at $B=38.05\,\mathrm{mT}$ as well as mixed states at (b)~$B=21.92\,\mathrm{mT}$
and (c)~$B=22.07\,\mathrm{mT}$.}
\label{fig:xrayimage}
\end{figure}
Under further increase of the magnetic induction we observe a transition
from the hexagonal pattern, shown in figure~\ref{fig:xrayimage}\,(a) to
a square one, as displayed in figure~\ref{fig:xrayimage}\,(d). This
transition has been previously observed by \citeasnoun{allais1985}
and \citeasnoun{abou2001}. The latter have investigated 
the role of penta--hepta defects for the mechanism of the transition.
Moreover, the evolution of the wavenumber was
measured for an adiabatic increase and a sudden jump of the magnetic induction.
However, this remarkable studies were limited to the planform of the patterns
and could not take into account the surface topography of the problem. Square patterns in MF have also been obtained in 
numerical investigations by \citeasnoun{boudouvis1987}. 

The transition between hexagonal and square planforms has been observed in other
experiments, like in B\'enard--Marangoni convection
\cite{thiele1998}, in nonlinear optics \cite{aumann2001} and in vibrated granular matter \cite{melo1995}.
In theory 
the competition between hexagons and squares has been studied for convection
\cite{malomed1990,bestehorn1996,bragard1998,herrero1994}.

The transition is especially interesting, 
because it is a smooth morphological one. \citeasnoun{kubstrup1996}, e.g., have
studied fronts between hexagons and squares in a generalized Swift--Hohenberg
model. They found pinning effects in domain walls separating different
symmetries, as suggested by \citeasnoun{pomeau1986}. These pinning effects are responsible for the static coexistence of
both patterns in an extended parameter range. Note that we have observed
wave-front pinning in the last paragraph as well within the context of ferrosolitons.
In the different context of ferromagnetism, pinning effects between different domains of magnetic ordering are known to cause hysteresis of the
order parameter  \cite{stoner1953,jiles1984}. What would be an appropriate order parameter for our
context, which is capable to unveil hysteresis?

First we test the local amplitude of the central peak and the wavenumber of the pattern. Next we focuse
on different order parameters tailored to this problem and adapted from various scientific fields. 

\subsection{Amplitude of the pattern in real space}
\label{sect:amplitude}
\begin{figure}
\setlength{\mathindent}{0pt}
\begin{minipage}{0.7\linewidth}
\includegraphics[width=\linewidth]{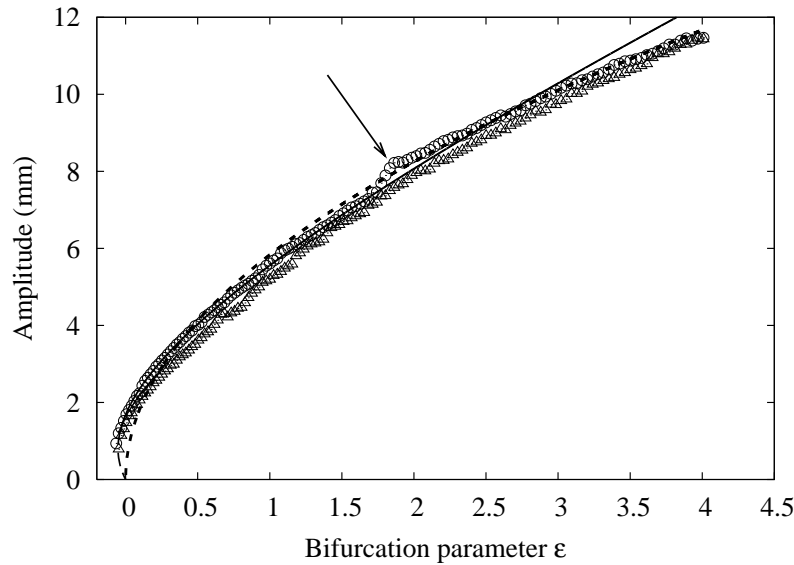}
\end{minipage}
\begin{minipage}{0.29\linewidth}
\caption{Amplitude of the nonlinear pattern in real space, extracted from one single peak in the
center of the dish. The triangles denote the amplitude for increasing induction, the circles for
decreasing induction. The solid line is a fit with~(\ref{eq:AE.hexagons}). The dashed line is a fit
of the upper branch for $\varepsilon>1.9$ with~(\ref{eq:AE.squares}). For the arrow see the body text.}
\label{fig:amplitudevsb}
\end{minipage}
\end{figure}
We recorded 500 images at different magnetic fields, raising the induction adiabatically from $0$ to
$38.1\,\mathrm{mT}$ and decreasing to zero afterwards. Figure~\ref{fig:amplitudevsb} shows the dependence of
the height of one single peak selected from the centre of the dish, and tracked thereafter, as a function of the applied
magnetic field. 
%During the whole process of increasing and decreasing the magnetic induction, the selected peak 
%floats around and comes as close as $0.5\,\mathrm{mm}$ to the centre of the vessel, while the
%longest  distance therefrom is $18.5\,\mathrm{mm}$. Considering that the container radius is ten
%times larger, the field inhomogeneity induced by the container edge should have no influence on its
%height on this scale.

At a first glance, the amplitude seems to be continuous in spite of the picture predicted by
\citeasnoun{gailitis1977} and \citeasnoun{friedrichs2001}. Obviously the height of the peak is only
slightly influenced by the geometry of the embedding pattern. A more careful comparison with 
the fit by the amplitude equation~(\ref{eq:AE.hexagons}) of the peak heights for decreasing
$\varepsilon$, however, reveals some deviations (see e.g. the  arrow in figure~\ref{fig:amplitudevsb}). 
In the range above the arrow the peak is situated inside a square domain and the amplitude can
therefore be fitted by~(\ref{eq:AE.squares}). 
Below it is embedded in a
hexagonal domain and hence, the fit by~(\ref{eq:AE.hexagons}) is a convincing description of the data. 
Because the shift of the amplitude is small, the latter is not a sensitive order parameter for
this transition. 

\subsection{Wavenumber of the pattern}
\label{sect:wavenumber}
\begin{figure}
\setlength{\mathindent}{0pt}
\begin{minipage}{0.7\linewidth}
\includegraphics[width=\linewidth]{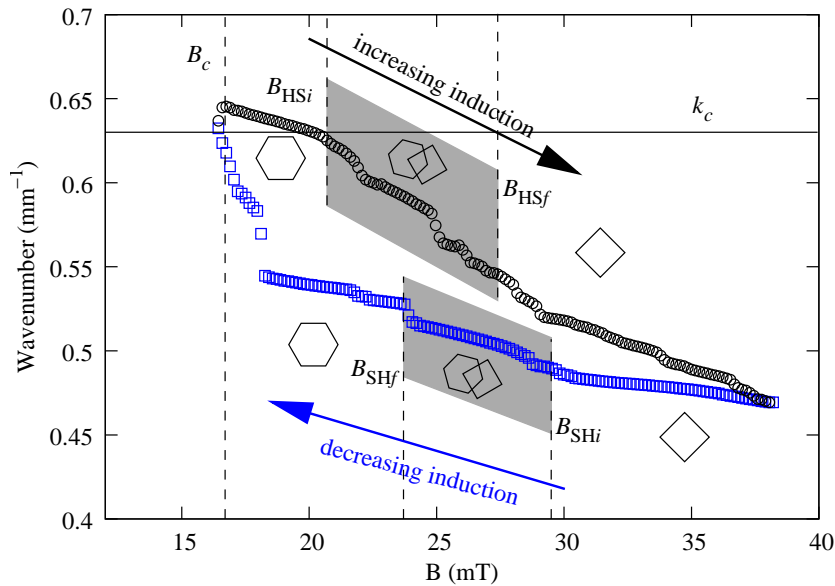}
\end{minipage}
\begin{minipage}{0.29\linewidth}
\caption{Wavenumber of the pattern as a function of the applied magnetic induction. Circles~(squares) denote
the values for increasing~(decreasing) induction. In the region shaded in grey we find coexistence
of hexagonal and square patterns. The horizontal line displays the capillary
wavenumber $k_c=\sqrt{\frac{\rho g}{\sigma}}=0.63\,\mathrm{mm^{-1}}$ for comparison.}
\label{fig:wavenumber}
\end{minipage}
\end{figure}
Another important measure, which discriminates between the two patterns is the wavenumber.
As shown by \citeasnoun{friedrichs2001}, the preferred wavenumber of squares is always less than
those of hexagons in the region of coexistence. 
Figure~\ref{fig:wavenumber} shows the wavenumber modulus as a function of the applied magnetic
induction determined from Fourier space.
We find a strong hysteresis, that goes alongside the different patterns.

When increasing the
magnetic induction, the surface remains flat up to the critical induction $B_c=16.7\,\mathrm{mT}$,
where a hexgonal array of peaks appears. Local quadratic ordering appears \emph{initially} at a threshold of
$\bhsi=20.7\,\mathrm{mT}$, and the whole container is \emph{finally} covered with a square pattern at
$\bhsf=27.4\,\mathrm{mT}$. Increasing the induction further, no new transition occurs until the
maximum induction $B=38.1\,\mathrm{mT}$ is reached. When decreasing the field, the first hexagons appear
already at $\bshi=29.5\,\mathrm{mT}$. Surprisingly, the comeback of the hexagons occurs at a higher
field than the transition therefrom, i.e. $\bshi>\bhsf$. The same holds true for the pure hexagonal
pattern, which comes back at  $\bshf=23.7\,\mathrm{mT}$, i.e. $\bshf>\bhsi$. We will refer to this
phenomenon as an 
inverse hysteresis, i.e. a \emph{proteresis} \cite{girard1989}. This observation is in agreement with \citeasnoun{abou2001}. The transition can
therefore not be explained by simple bistability. The wavenumber is
continuous during the transition, but it has a dramatic increase between \bshf\  and
$B_c$ when decreasing the field, where it quickly relaxes to the initial value.  
The general tendency, that the wavenumber decreases with increasing field agrees with theoretical
predictions by \citeasnoun{friedrichs2001}, whereas the hysteresis effect has not been predicted. 

%The wavenumber modulus is averaged over the whole pattern, whereas the amplitude is provided
%only on the basis of local information. This might be the reason that the hysteresis is not prominent 
%in the amplitude scaling. 

Neither the amplitude nor the wavenumber can clearly distinguish between square and
hexagonal patterns. The amplitude is provided only on the basis of local information and is
therefore only slightly influenced by the geometry of the pattern. The wavenumber comprises global information and
depends strongly on the pattern, but from its magnitude alone it is not possible to decide about the
actual geometry. In conjunction with the distance between two adjacent peaks, however, this is possible as shown below. 

\subsection{Analysing the peak--to--peak distance}
\label{sect:peaktopeak}
\begin{figure}
\centering
(a)\includegraphics[width=0.4\linewidth]{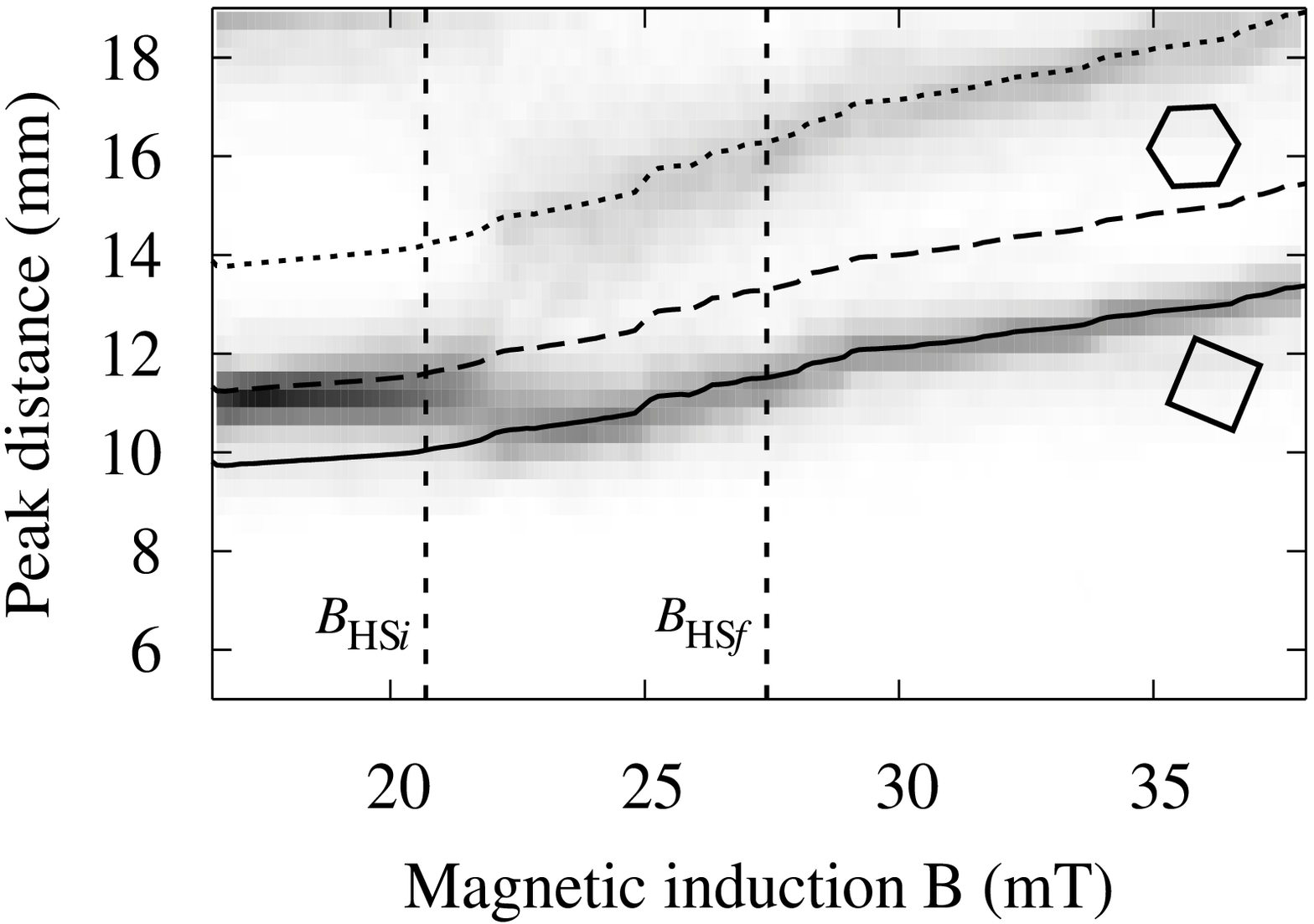}
(b)\includegraphics[width=0.4\linewidth]{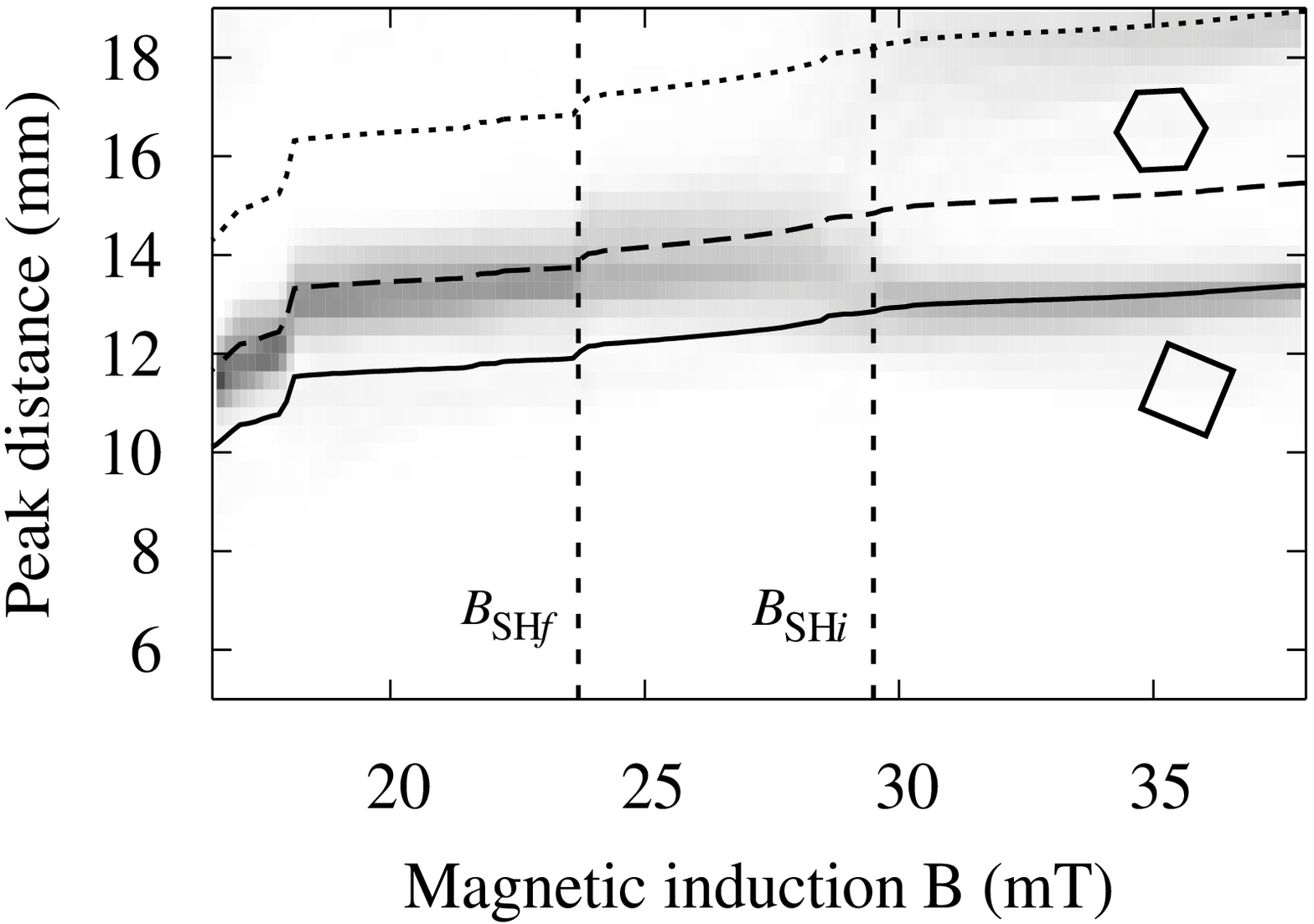}

\caption{Histogram of the peak--to--peak distance for increasing~(a) and decreasing
induction~(b).  
A high count is represented by a dark shade of gray. The lines display the theoretical distance for
regular hexagons~(dashed), squares~(solid), and the next-to-nearest neighbours of squares~(dotted) 
derived from the measured wavenumber.  }
\label{fig:peaktopeak}
\end{figure}
The relation between the wavelength and the peak--to--peak distance of nearest neighbours is
different for hexagons and squares: The peaks on the hexagonal grid are wider spaced by a factor of
$\sqrt{3/2}$. We therefore plot a histogram of the peak--to--peak distance in
figure~\ref{fig:peaktopeak} versus $B$. Thus we can compare the distance distribution to the wavelength,
which is plotted on top. 

In spite of the visual observation of the transition (marked by the dashed vertical lines in
figure~\ref{fig:peaktopeak}\,a), the peak--to--peak distance makes a rather sudden jump from the
hexagonal to the square state, when the induction goes up. When decreasing the field
(figure~\ref{fig:peaktopeak}\,b), the
distribution of the peak--to--peak distances broadens in the region of coexistence.

In contrast to the plane, averaged wavelength, its combination with the peak--to--peak distance  is able to discriminate
between trigonal and square symmetry. This is due to a combination of information from real space
and Fourier space. We will now apply this promising concept to the amplitude and its Fourier domain
analogon. 
\subsection{Fourier-domain based correlation}
\begin{figure}
\centering
(a)\includegraphics[width=0.2\linewidth]{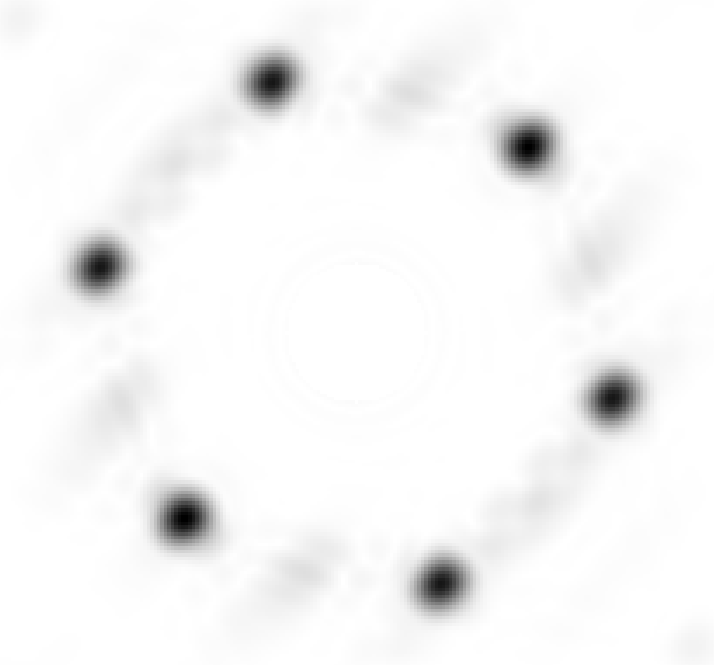}
(b)\includegraphics[width=0.2\linewidth]{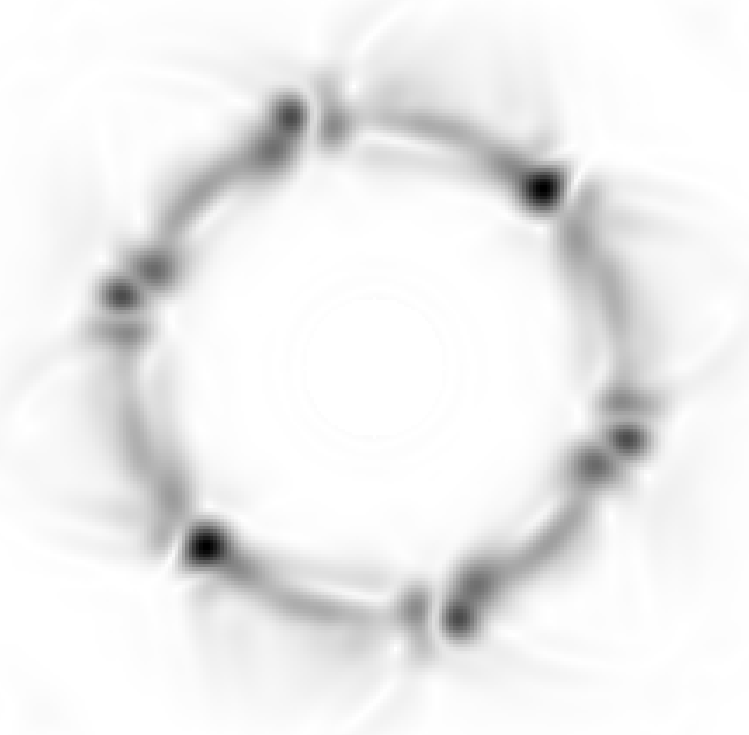}
(c)\includegraphics[width=0.2\linewidth]{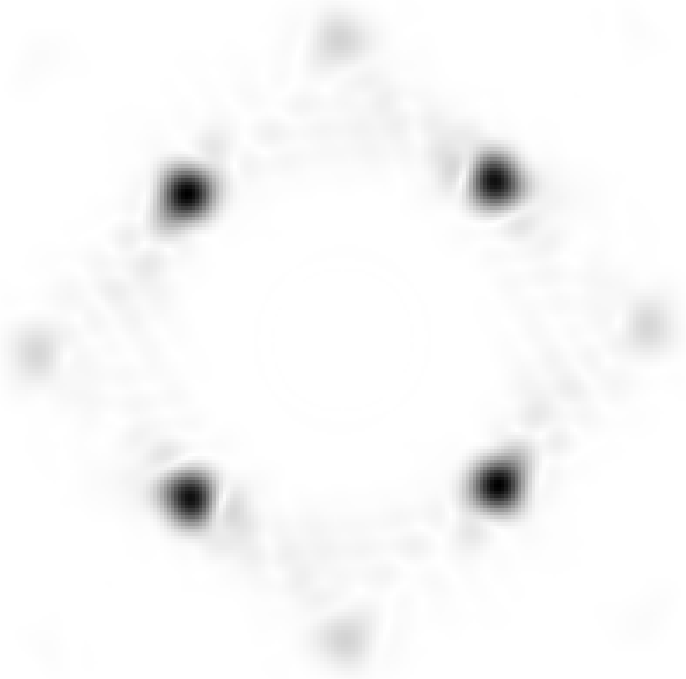}

\caption{Fourier transform of the pattern at three different magnetic inductions with increasing
induction. (a)~$B=19.78\,\mathrm{mT}$ (b)~$B=22.07\,\mathrm{mT}$ (c)~$B=28.92\,\mathrm{mT}$ } 
\label{fig:fourierdomain}
\end{figure}

The symmetry of the pattern is reflected in the Fourier space, as shown in 
figure~\ref{fig:fourierdomain}.  
Both the hexagonal and square pattern produce clear individual peaks~(a,c), while the transform of a mixture
of both patterns results in a ring~(b). 
With an angular correlation function, following
\citeasnoun{millan1996}, it is possible to discriminate between the amplitude contribution of each pattern: Let
$\mathcal{F}(k_x, k_y)$ be the discrete Fourier transform of the surface relief of the pattern
\begin{equation}
\mathcal F (k_x, k_y) = \sum_{x=0}^{m-1} \sum_{y=0}^{n-1} 
                       h(x,y) \exp\left(-2\pi i (k_x x/m + k_y y/n)\right)
\label{eq:intensity}
\end{equation}
where $h(x,y)$ is the height of the fluid surface at the point $(x,y)$ and $m,
n$ are the number of pixels in $x$ and $y$ direction, respectively. Then
$\mathcal{A}=\left|\mathcal F(k_x, k_y)\right|$ denotes the corresponding
amplitude.  To suppress artifacts of the Fourier transform coming from the
boundary, we apply a cylindrically symmetric Hamming window of the width $b$
with the weight function
\begin{equation}
 w(x,y) = \cases{
  \left(0.54+0.46 \cos\left(  \frac{\pi \sqrt{x^2+y^2}}{b}\right)\right)^2 & for $x^2+y^2
  \leq b^2$ \\
   0 & else.
   }
   \label{eq:windowfunction}
\end{equation}
Then we define the Fourier angular correlation $P_\alpha$ to be
\begin{equation}
P_\alpha = \sum_{\vect{k}} \mathcal{A}(\vect{k})\mathcal{A}(R_\alpha\vect{k}),
\label{eq:correlation}
\end{equation}
where $R_\alpha$ is the matrix for rotation by the angle $\alpha$. Technically, the rotation by an
arbitrary angle is done using \possessivecite{paeth1986} algorithm. 

\begin{figure}
\setlength{\mathindent}{0pt}
\begin{minipage}{0.5\linewidth}
\includegraphics[width=\linewidth]{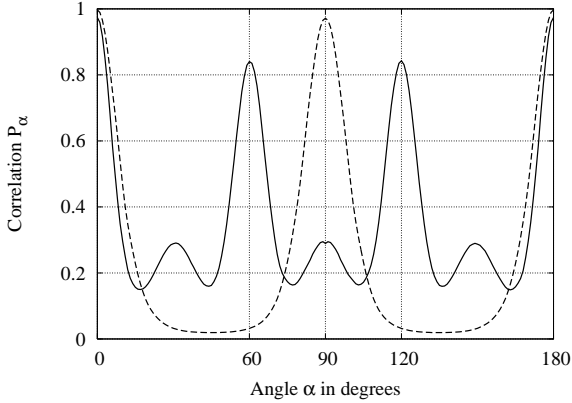}
\end{minipage}
\begin{minipage}{0.4\linewidth}
\caption{Angular correlation of the pattern as a function of the angle. Solid line: Pure hexagonal
pattern with a grain boundary at $B=17.5\,\mathrm{mT}$. Dashed line: Pure square pattern at
$B=38.1\,\mathrm{mT}$.}
\label{fig:fullangularcorr}
\end{minipage}
\end{figure}
This correlation function is displayed in figure~\ref{fig:fullangularcorr} for two different patterns. 
For a perfect hexagonal lattice the Fourier transform consists of a series of delta peaks, 
that is invariant under a rotation by
$60$ degrees. Consequently, the above defined correlation function would be zero for any angle
$\alpha$ that is not an integer multiple of $60$ degrees. The same argument applies to a perfect
square lattice that is invariant under a rotation by $90$ degrees. Therefore the hexagonal pattern
at  $B=17.5\,\mathrm{mT}$ manifests itself in major peaks at $60$ and $120$ degrees, while the
square pattern at $B=38.1\,\mathrm{mT}$ yields a strong peak at $90$ degrees. 
Further, the correlation
function will be proportional to the square of the amplitude of the corresponding lattice. Using the
above defined window function~(\ref{eq:windowfunction}), the amplitudes $A_6, A_4$ of a square and
hexagonal sinusoidal lattice in real spacen relate to the Fourier angular correlation function like
\begin{equation}
 A_6 = 4.5\sqrt{\frac{P_{60^\circ}}{1.5\,m n \xi }} \quad\mbox{and}\quad
 A_4 = 4\sqrt{\frac{P_{90^\circ}}{{m n \xi }}}.
\label{eq:correlationamplitude}
\end{equation}
Here 
\begin{equation}
\xi = \int w^2(x, y) \mathrm{d}x\mathrm{d}y =0.1032300545\,\pi b^2
\end{equation}
is a normalization factor for the window function. 

\begin{figure}
\centering
(a)\includegraphics[width=0.45\linewidth]{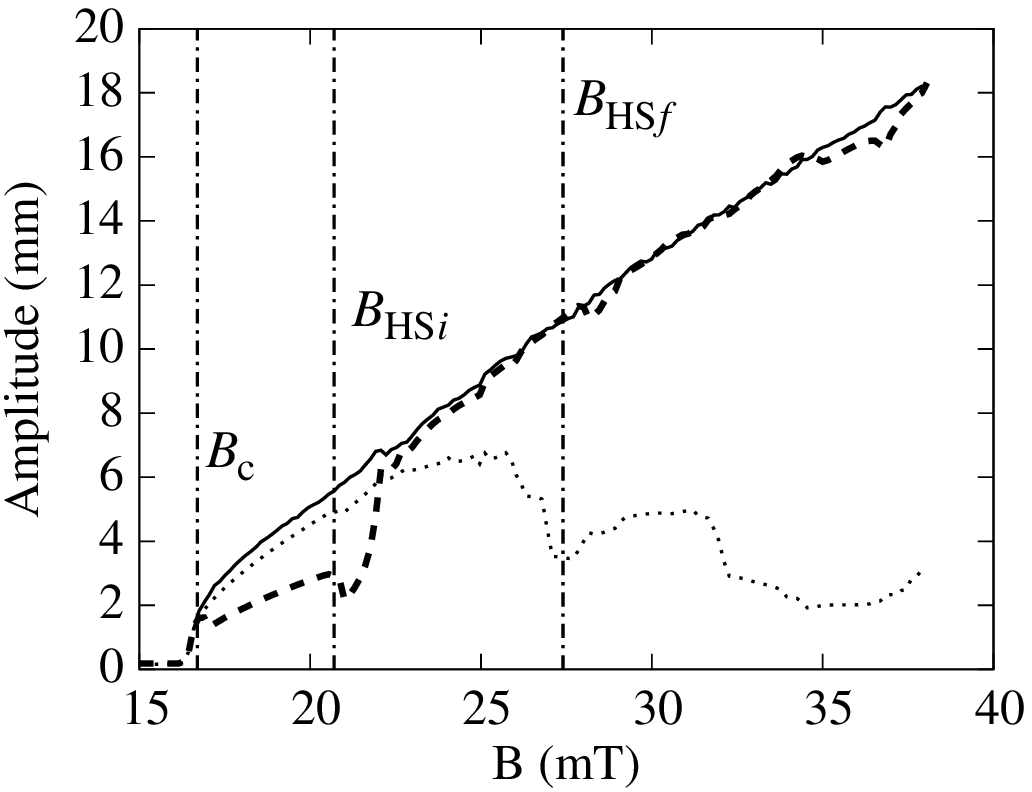}
(b)\includegraphics[width=0.45\linewidth]{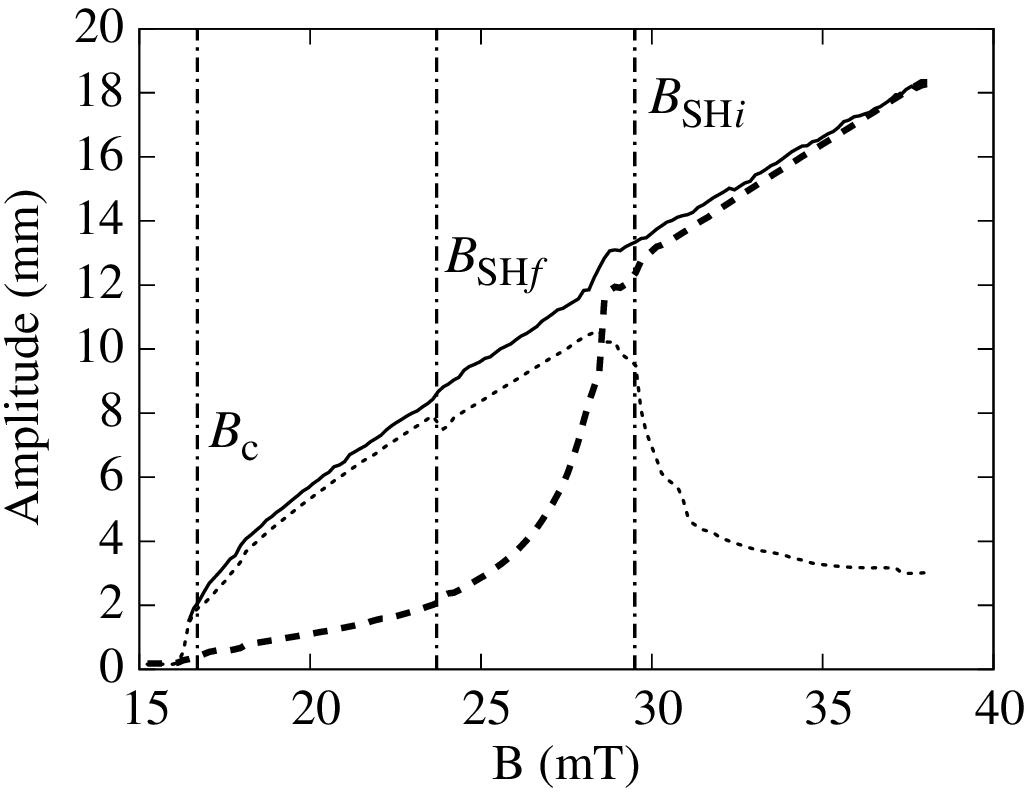}

\caption{Amplitude of the central peak $A$~(solid) and Fourier domain correlation amplitudes $A_4$~(dotted) and
$A_6$~(dashed) of the pattern for
increasing~(a) and decreasing~(b) magnetic induction.}
\label{fig:correlationamplitude}
\end{figure}
Figure~\ref{fig:correlationamplitude} shows the amplitude $A$ in real space, as discussed in
section~\ref{sect:amplitude}, together with $A_6$ and $A_4$. The amplitude
of the single peak in the centre is cum grano salis an upper bound of $A_6$ and
$A_4$. However, the smaller of both has an unnegligible value even for ranges, where
the pattern appears to be homogeneous to the naked eye. For example, the hexagonal range $[B_c,\bhsi]$ in
figure~\ref{fig:correlationamplitude}\,(a) 
gives still a finite $A_4$. This is due to two differently oriented hexagonal patterns, which are tilted
by an angle of 30 degrees. They are separated by a grain-boundary made of
penta--hepta defects (see figure~\ref{fig:xrayimage}\,a). 
The tilted smaller patch manifests itself as the intermediate structure between the major peaks in Fourier
space in figure~\ref{fig:fourierdomain}\,(a). This leads to a positive intercorrelation between those two patterns in Fourier space at 90 degrees. 

When increasing the field, this grain boundary moves to the border immediately before $B=\bhsi$ and
therefore $A_4$ goes down. Right after this the transition to squares at $B=\bhsi$ leads to
an increase of $A_4$, which continues to follow the amplitude of the central peak,
while $A_6$ goes down, indicating that the hexagonal pattern vanishes.
However, even though at $B=\bhsf$ the whole dish is apparently covered with squares, $A_6$ does
not decay to zero. This is due to the noisefloor, that always leads to a finite correlation
in any direction, as illustrated  by the dashed line in figure~\ref{fig:fullangularcorr}. 

The character of the transition between trigonal and square symmetry is a smooth one, which becomes
clear from figure~\ref{fig:correlationamplitude}\,(b). Both order parameters $A_4$ and $A_6$ are
continuous at the transition point \bhsi. This is an effect of smoothing 
due to the stepwise transformation of small blocks. Likewise the magnetization curve of a ferromagnet 
appears to be smooth, although the individual domains change their magnetization discontinuously. 

%Thus we have a second order transition, whereas the
%transition from the flat state to hexagons is of first order. 

In the next section we take a closer look at the mechanism for the smooth transition. For that
purpose we inspect Voronoi diagrams.
\subsection{Voronoi diagram for local information in real space}
\label{sect:voronoi}
\begin{figure}
\centering
\parbox{0.24\linewidth}{%
\begin{center}
\includegraphics[width=\linewidth]{figure12_a.eps}\\
(a)
\end{center}}\hfill
\parbox{0.24\linewidth}{%
\begin{center}
\includegraphics[width=\linewidth]{figure12_b.eps}\\
(b)
\end{center}}\hfill
\parbox{0.24\linewidth}{%
\begin{center}
\includegraphics[width=\linewidth]{figure12_c.eps}\\
(c)
\end{center}}\hfill
\parbox{0.24\linewidth}{%
\begin{center}
\includegraphics[width=\linewidth]{figure12_d.eps}\\
(d)
\end{center}}\\
\includegraphics[width=0.8\linewidth]{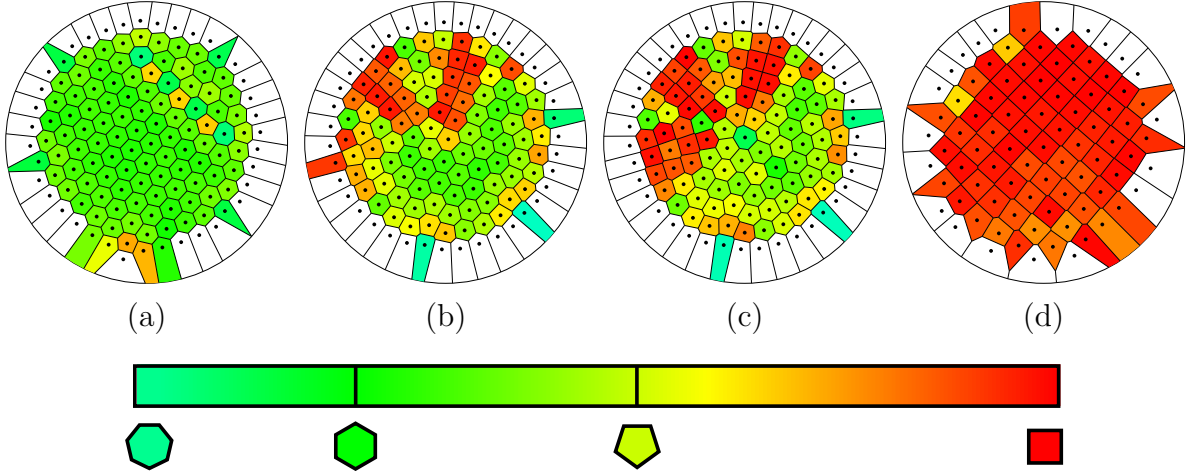}

\caption{Coloured Voronoi diagram over the centre of the peaks for (a)~$B=19.78\,\mathrm{mT}$,
(b)~$B=21.92\,\mathrm{mT}$, (c)~$B=22.07\,\mathrm{mT}$, and (d)~$B=38.05\,\mathrm{mT}$. The bar
below indicates the colour code,
which is derived from the maximum central angle $\alpha$ of each cell of the pattern. 
The colour code hue value in the HSB colour space is set to $h=1-{\nicefrac{2\alpha}{\pi}}$ with full
brightness and saturation.
}
\label{fig:voronoi}
\end{figure}
A classical tool for the analysis of nearest neighbours of a set of points, called sites, is the
Voronoi diagram~\cite{fortune1995}. Four Voronoi diagrams for increasing magnetic induction are
shown in figure~\ref{fig:voronoi}.
The two-dimensional Voronoi tesselation over a discrete set of sites 
\[S=\left\{ \vect{x_i}\right\}, \vect{x_i} \in \mathbb{R}^2\] 
is defined in terms of nearest neighbours: the Voronoi cell $V_\vect{x_i}$ corresponding to the site
$\vect{x_i}$ is the set of points, that is closer to $\vect{x_i}$ than to any other site:
\begin{equation}
V_{\vect x_i} = \left\{ \vect x \in \mathbb{R}^2, \quad \left| \vect x - \vect
x_i \right| <  \left| \vect x - \vect x_j\right| \quad \forall j,\, j \neq i \right\} 
\end{equation}
The Voronoi cells are the interior of convex polygons which tesselate the entire space
$\mathbb{R}^2$. The nearest neighbouring sites of any given site $\vect{x_i}$ are then defined by all
$\vect{x_j}$, where the Voronoi cell $V_\vect{x_j}$ shares a common edge with $V_\vect{x_i}$. Note
that these ``nearest neighbours'' don't have the same distance from $\vect{x_i}$ in general. This is 
only true for perfectly regular lattices.

We extract the position of the peaks with subpixel accuracy by fitting a paraboloid to the centre of
each peak. From these coordinates we construct the Voronoi diagram. Since the Voronoi diagram of a
regular hexagonal or square lattice is a regular hexagonal or square tesselation, where the sites
are the centre of the cells, one might expect that the number of edges serves as a criterion of the
local ordering. Unfortunately, a real quadrilateral is singular in the Voronoi diagram. An
infinitesimal slight distortion of a square lattice of sites results in mostly hexagonal Voronoi cells
with two tiny edges, that look like squares. But there is a number of 
metric quantities listed by  \citeasnoun{thiele1998}, namely angle, cell perimeter and cell area, which 
are not affected and can therefore be employed.  The diagrams in figure~\ref{fig:voronoi} are
colour coded with the maximum central angle of each cell.

This method can visualize the mechanism for the smooth transition. In
figure~\ref{fig:voronoi}\,(a) almost all Voronoi cells are hexagonal (green), apart from a line of
penta--hepta defects (dark green and yellow). Upon increase of $B$, this pattern remains stable up to
\bhsi. For $B$ slightly above \bhsi, two domains of square cells have been formed
(figure~\ref{fig:voronoi}\,b). The invasion of the hexagonal pattern by the square one takes place
domainwise. This is corroborated when looking at (c), where another patch of squares has emerged. 
Obviously the domain-like transformation of
the pattern is responsible for the overall smooth transition. A sudden transform into a pure square
array is probably hindered by pinning of the domainfronts. The final square state is shown in
figure ~\ref{fig:voronoi}\,(d). 

In the next section, we employ the remaining metric parameters for
quantitative analysis. 
\subsection{Cell perimeter and area}
\begin{figure}
\centering
(a)\includegraphics[width=0.45\linewidth]{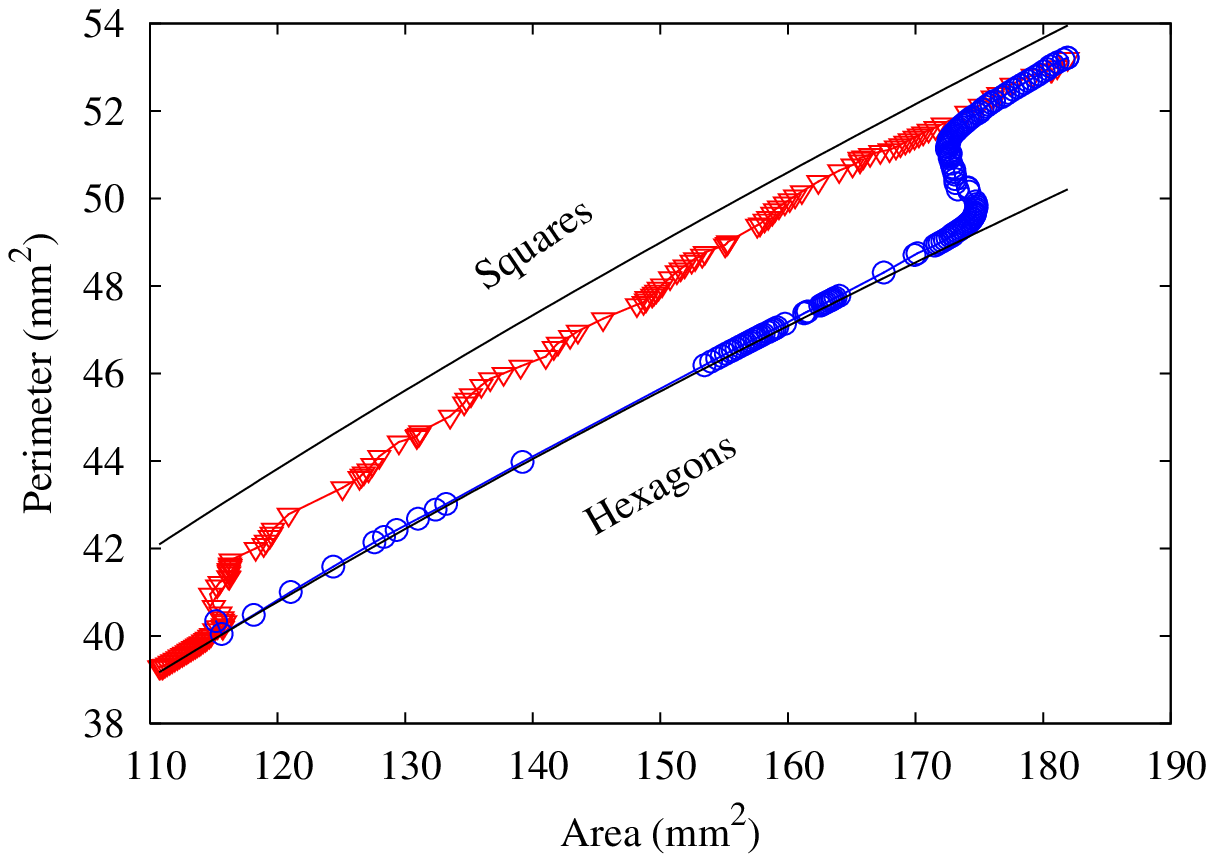}
(b)\includegraphics[width=0.45\linewidth]{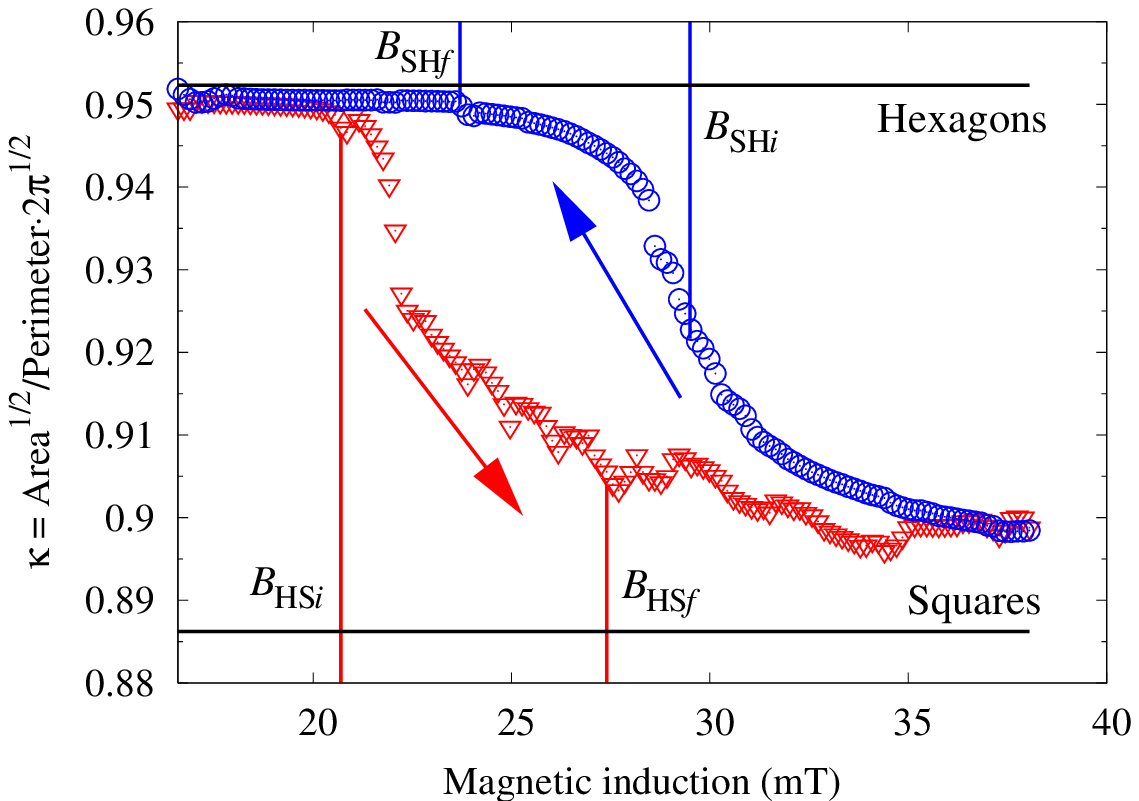}

\caption{Hysteretic transition from hexagons to squares is shown (a) in the perimeter--area space,
and (b) in the plot of the circularity $\kappa$ versus the magnetic induction. Triangles~(circles) denote increasing~(decreasing) magnetic induction. } 
\label{fig:sgpn}
\end{figure}

According to \citeasnoun{thiele1998}, the perimeter--area-ratio can be utilized to distinguish
between hexagonal and square symmetry. The relations of the perimeter $\Pi$ and the area $a$ of regular
planforms differ for hexagons and squares; as a function of the wavelength $\lambda$ they fulfill the
following equations:
\begin{equation}
\Pi_6 = 4 \lambda,  \quad \Pi_4 = 4\lambda, \quad a_6 = \frac 2 3 \sqrt{3} \lambda^2,
   \quad \mbox{and}\quad a_4 = \lambda^2
\end{equation}
Thus, a plot of the average perimeter versus the average cell area leads to a square root with
different factors:
\begin{equation}
\Pi_6 (a_6) = 4\sqrt{\frac{\sqrt{3}}{2}}\sqrt{a_6} \quad\mbox{\textrm{and}}\quad
\Pi_4 (a_4) = 4 \sqrt{a_4}
\end{equation}

Figure~\ref{fig:sgpn}\,(a) shows this plot for the experimental data, which approach the theoretical relations for
regular patterns in a hysteretic manner. 
However, in this diagram the dependence on the control parameter $B$
is lost. We therefore define the \emph{circularity $\kappa$}, a normalized area--perimeter ratio, by
\begin{equation}
\kappa:=\frac{\sqrt{\left<a\right>}}{\left<\Pi\right>}\cdot 2 \sqrt{\pi},
\end{equation}
which becomes $1$ for a perfect circle and is less for polygons. 
It is shown in figure~\ref{fig:sgpn}\,(b) as a function of the magnetic induction.
A regular tesselation for hexagons results in
$\kappa_6=\sqrt{\frac{\pi}{6} \sqrt{3}}\approx0.952$ and $\kappa_4=\frac{\sqrt{\pi}}{2}\approx0.886$, respectively. 

This order parameter clearly shows that the hexagonal ordering is nearly perfect, because the
experimental data meet the theoretical limit $\kappa_6$ quite well. For increasing $B$, $\kappa$ decays on a
jerky path, indicating a blockwise transformation from trigonal to square symmetry -- a macroscopic
analogon to the Barkhausen effect. Obviously, the minimal value $\kappa_4$ is not reached, probably
due to the instability of squares in the Voronoi tesselation. 
Under decrease of $B$, $\kappa$ follows a different, smoother path. All in all, the circularity shows a
broad proteresis, which behaves inverse to the common hysteresis in
that way, that it shows no retardation, but an advanced behaviour. 

The visualization in section~\ref{sect:voronoi} uses the maximum central angle of the cells. Now we
want to propose a method that makes use of all central angles of the Voronoi cells. 
\subsection{Angular correlation function}
\begin{figure}
\centering
(a)\includegraphics[width=0.45\linewidth]{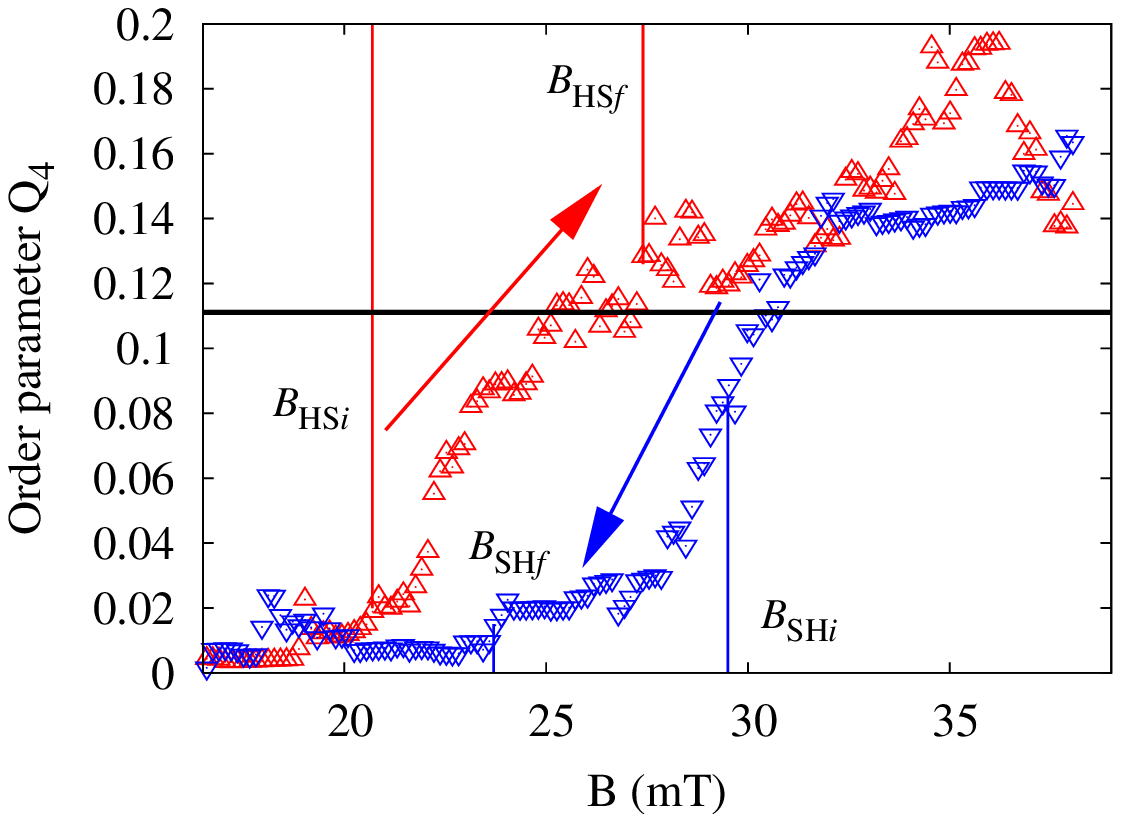}
(b)\includegraphics[width=0.45\linewidth]{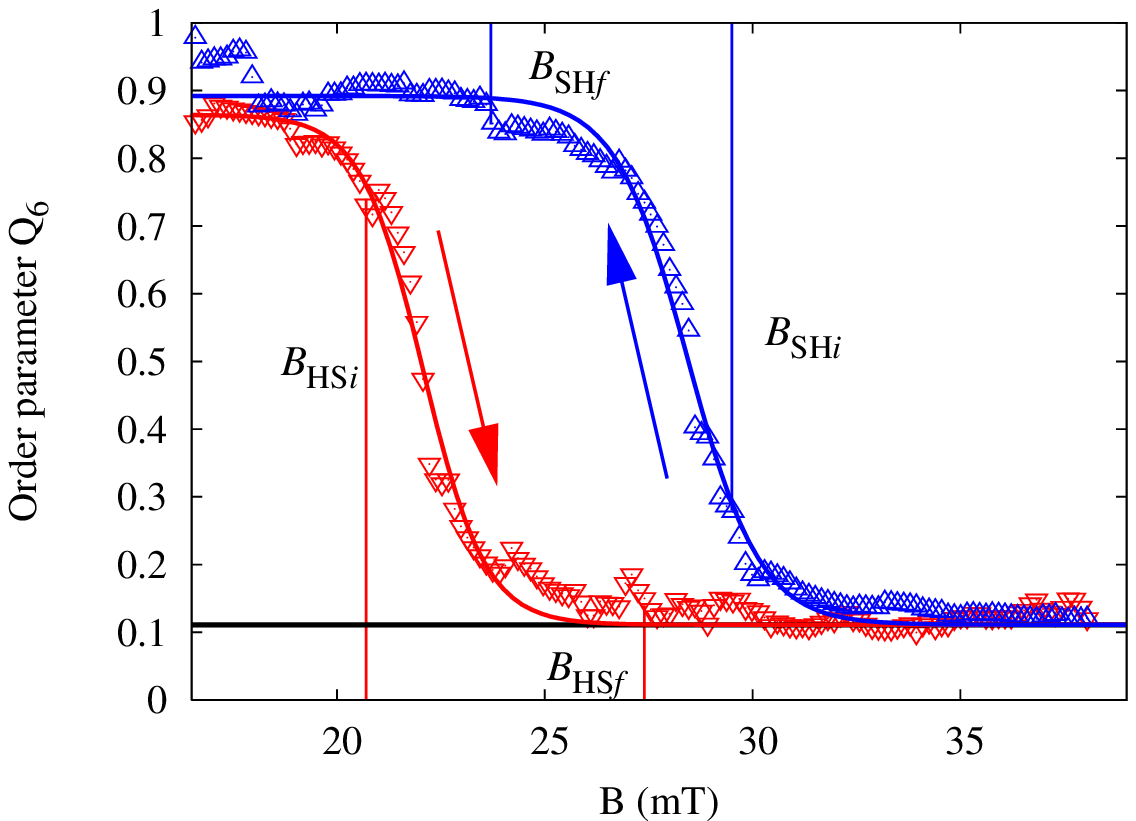}

\caption{Angular correlation functions $Q_4$~(a) and $Q_6$~(b) in real space. The solid line in (b)
marks a fit with the logistic function. }
\label{fig:qangularcorr}
\end{figure}
Inspired by 
\citeasnoun{steinhardt1983}, who introduce a local order parameter in three dimensions, we propose
the local angular correlation $Q_m$ as a transformation of the central angles:
\begin{equation}
Q_m = \left| \frac{1}{N} \sum_{i=1}^N e^{im\phi_i} \right|^2
\end{equation}
where $\phi_i$ is the angle enclosed by an arbitrarily chosen axis and the line through the selected
peak and its $i$-th nearest neighbour. For regular polygons with $N$ sides, this parameter gives $1$
exactly when $m$ is an integer divisor of $N$ and $0$ otherwise. For irregular polygons, the result may be
any value in the range $\left[0\ldots1\right]$. 
We apply the parameters $Q_4$ and $Q_6$ to our
experimental data and average over all cells. Figure~\ref{fig:qangularcorr}\,(a) shows the hysteretic
behaviour of $Q_4$. Though it is a normalized measure, it approaches by no means $1$. A possible
explanation is that a large number of the squares are indeed degenerated hexagons, i.e. two corners are cut off at $45^\circ$. If all
squares would be degenerated in that way, the maximum would be $Q_4=\frac{1}{9}$, represented by the
solid horizontal line. 

The complementary measure $Q_6$, as shown in figure~\ref{fig:qangularcorr}\,(b), does not suffer from
these problems. It shows a smooth, nearly ideal hysteretic cycle, which can be fitted by two 
logistic functions. Most importantly, $Q_6$ clearly reflects the visual impression of the pattern,
as revealed by comparison with the thresholds.

\section{Summary and Conclusion}
The topic of this article is twofold. First, we reported on
measurements of the hexagonal surface topography, evolving after the
flat surface becomes unstable due to a transcritical bifurcation. By
means of radioscopy we have access to the surface reliefs. The
scaling of the amplitudes obtained therefrom
is in agreement with the equations deduced by
\citeasnoun{friedrichs2001}. Moreover, comparing the amplitudes
and peak profiles from the measurements with the results obtained by \citeasnoun{matthies2005} via FEM numerics matches
without further fitting parameters. Eventually, we
can generate \emph{static} localized states (ferrosolitons) in the bistable
regime, where hexagons coexist with the flat reference state, by
perturbing the surface locally.

The second part is devoted to new experimental results regarding
the hexagon--square transition. Its characterization is difficult, because
trigonal and quadrilateral symmetry coexist in an extended parameter range. 
From the recorded surface reliefs we have
extracted a set of order parameters. The classical measurands, namely the amplitude and
the wavenumber, show a hysteretic behaviour, but are not specific enough to catch
the geometry. The combination of methods from the real and inverse space, like the
peak--to--peak distance and the Fourier domain based correlation, uncovers a
smooth transition between both patterns. The Voronoi diagrams reveal a domainwise
transformation, suggesting a pinning mechanism. We have tailored two further
measures, namely the circularity $\kappa$ and the angular correlation $Q$, which
can map the complex transition in agreement with the visual impression. These
show an inverse (advanced) hysteresis, called proteresis in pharmacodynamics \cite{girard1989}. 
These measures may be applied to numerical calculations as well as to other
experiments, providing a basis for
comparison. 

Both hysteretic transitions originate from the bistability of two patterns. In the first case, the
energy barrier between the flat state and the hexagons is huge. Only by applying a local
perturbation we can observe both states side by side in one container. The ferrosolitons are
hindered to penetrate into the flat surface by pinning of the wavefronts. In the second case, the
energy barrier between hexagonal and square structures appears to be lower. The pinning of the domainfronts can
be overcome under variation of the control parameter, which results in a blockwise transformation.
By applying noise, we expect that the first transition could be smoothed out in a similiar manner. 
%changed to second order. 

\section*{Acknowledgements}
We wish to thank R. Friedrichs for putting his data (figure~\ref{fig:intro}\,b) to our disposal.
Moreover, we would like to thank I.V. Barashenkov, A. Lange, O. Lavrova, G. Matthies and L. Tobiska for the fruitful cooperation
and A. G\"otzendorfer, A. Engel,  P. Leiderer,  K. Morozov and  
M. Shliomis for helpful discussions and advice. The
authors gratefully acknowledge that the reported experiments have
been funded within the priority program SPP1104 under grant Ri\,1054/1.

\section*{References}
\bibliographystyle{jphysicsB}
\bibliography{alles}

\end{document}